\documentclass[twocolumn,showpacs,preprintnumbers,amsmath,amssymb,superscriptaddress]{revtex4-1}

\usepackage{graphicx,footnote}
\usepackage{epstopdf} 
\usepackage{enumerate}
\usepackage[format=plain,singlelinecheck=off]{caption}
\usepackage{subcaption}
\usepackage{ngerman}
\usepackage[utf8]{inputenc}
\usepackage{amsmath}
\usepackage[stable]{footmisc}
\usepackage{dsfont}
\usepackage{textcomp}
\usepackage{listliketab}
\usepackage{tabularx}
\usepackage{multirow}
\usepackage{color}
\usepackage{tabto}
\usepackage{stmaryrd}
\usepackage{amssymb}
\usepackage{here}
\usepackage{footnote}
\usepackage{pstricks,pstricks-add}
\usepackage{tikz}
\usepackage{pst-plot}
\usepackage{color}

\makeatletter 
\g@addto@macro\normalsize{% 
\setlength{\abovedisplayskip}{5pt} 
\setlength{\belowdisplayskip}{10pt}} 
\makeatother

\begin{document}

\title{Dynamics of test particles in the five-dimensional G\"{o}del spacetime}

\author{Kevin Eickhoff}
\email{kevin.eickhoff@uni-oldenburg.de}
\affiliation{Institut f\"{u}r Physik, Universit\"{a}t Oldenburg, 26111 Oldenburg, Germany}

\author{Stephan Reimers}
\email{stephan.reimers@uni-oldenburg.de}
\affiliation{Institut f\"{u}r Physik, Universit\"{a}t Oldenburg, 26111 Oldenburg, Germany}

\date{\today}

\begin{abstract}
	\noindent
	We derive the complete set of geodesic equations for massive and massless, charged test particles of a five-dimensional, rotating and charged solution of the Einstein-Maxwell-Chern-Simons field equations in five-dimensional minimal gauged supergravity and present their analytical solutions. We study the polar and radial motion, depending on the spacetime and test particle parameters, and characterize the test particle motion qualitatively by the means of parametric plots and effective potentials. We use the analytical solutions in order to visualize the test particle motion by three-dimensional plots.
\end{abstract}

\maketitle

\section{Introduction}

The G\"{o}del metric is an exact regular solution of the Einstein field equations in the presence of a negative cosmological constant, which was published in 1951 by Kurt G\"{o}del \cite{Godel:1949ga} as a gift to Einstein's 70th Birthday. It describes a homogeneous pressureless mass distribution.\\
This solution respresents the best known example of an universe model with causality violation (e.g. the existence of closed timelike curves). Closed timelike curves are also found in the van Stockum spacetime of a rotating dust cylinder \cite{vanStockum:1937zz}, the Kerr spacetime \cite{Kerr:1963ud} and the Gott spacetime of two cosmic strings \cite{Gott:1990zr}.\\
Additionally, the G\"{o}del spacetime was the first solution of the Einstein field equations that modelled a globally rotating universe, demonstrating that Mach's principle \cite{Bondi:1996md, Barbour:1995iu} is not fully incorporated in the theory of general relativity.\\ 
Although not serving as a viable model of our universe, since it does not include any expansion as required by Hubble’s law \cite{Hubble:1929ig}, G\"{o}del's solution gave rise to general questions of causality and global properties of relativistic spacetimes, which culminated in the postulation of the chronology protection conjecture by Stephen Hawking \cite{Hawking:1991nk}.\\ 
As a promising candidate for a quantum theory of gravity, string theory generated a growing interest in higher-dimensional solutions, since it requires extra dimensions of spacetime for its mathematical consistency. The higher-dimensional generalization of the Schwarzschild spacetime has been found in 1963 by F. R. Tangherlini \cite{Tangherlini:1963bw}. In 1986, R. Myers and M. Perry generalized the Kerr solution to higher dimensions \cite{Myers:1986un}. Further generalizations include the general Kerr-de Sitter and Kerr-NUT-AdS metrics in all higher dimensions \cite{Gibbons:2004uw,Chen:2006xh}. Remarkably, five-dimensional, stationary vacuum black holes are not unique. Besides the Myers-Perry solution, a five-dimensional rotating black ring solution with the same angular momenta and mass but a non-spherical event horizon topology have been found \cite{Emparan:2001wn}.\\
However, neither the four-dimensional Kerr-Newman nor the G\"{o}del solution of Einstein's field equation could be generalized to higher dimensions, yet. Nevertheless, related solutions of both spacetimes were found for the Einstein-Maxwell-Chern-Simons (EMCS) equations of motion in the five-dimensional minimal gauged supergravity \cite{Gauntlett:2002nw, Chong:2005hr}. The maximally supersymmetric G\"{o}del analogue shares most of the peculiar features of its four-dimensional counterpart (see e.g. \cite{Buser:2013uaa}).  %\cite{Gimon:2003ms, Gauntlett:2004cm, Kunduri:2005vc}\\
The test particle motion, governed by the geodesic equations, is a valuable tool in order to gain insight into the fundamental properties of a spacetime. Especially, exact solutions of the geodesic equations can be used to calculate spacetime observables to arbitrary accuracy. Further interest into geodesics in anti-de Sitter spacetimes arises in the context of string theory and the AdS/CFT correspondence \cite{Maldacena:1997re}. The geodesic equations of the four-dimensional G\"{o}del spacetime have been investigated in \cite{Grave:2009zz}. The geodesic equations of the five-dimensional Kerr-Newman analogue of the EMCS equations of motion were solved analytically in \cite{Reimers:2016czc}.\\
The separability of the geodesic equations for non-extremal rotating charged black holes in the Gödel universe of the five-dimensional minimal supergravity theory (see \cite{Gimon:2003ms}) was shown in \cite{Wu:2007gg}.\\
In this paper, we want to explore the dynamics of test particles coupled to the U(1) field of the five-dimensional G\"{o}del spacetime and solve the geodesic equations analytically. In Sec. II, we will present the basic features of this spacetime and derive the geodesic equation by solving the Hamilton-Jacobi equation. Sec. III contains a qualitative discussion and a complete characterization of the test particle dynamics, especially the radial effective potentials are introduced. Sec. IV is dedicated to the analytical solutions of the equations of motions obtained in Sec. II, which will be used in Sec V, in order to illustrate some three-dimensional representations of the related orbits.

\section{The five-dimensional G\"{o}del universe}

We will briefly recall the basic properties of the five-dimensional G\"{o}del spacetime and derive the geodesic equations describing the  motion of massive and massless test particles.

\subsection{Metric}

The bosonic part of the minimal supergravity theory in 4+1 dimensions consists of a metric and a one-form gauge field obeying the Einstein-Maxwell-Chern-Simons (EMCS) equations of motion \cite{Gauntlett:2002nw}

\begin{align}
\label{eq:EMCS}
R_{\mu \nu} - \frac{1}{2} g_{\mu \nu} R = 2 \left(F_{\mu \alpha} F_\nu{}^\alpha - \frac{1}{4} g_{\mu \nu} F_{\rho \sigma} F^{\rho \sigma} \right),\\
\nabla_\mu \left(F^{\mu \nu} + \frac{1}{\sqrt{3} \sqrt{-g}} \epsilon^{\mu \nu \lambda \rho \sigma} A_\lambda F_{\rho \sigma}\right) = 0,
\end{align}

where $F_{\mu \nu} = \partial_\mu A_\nu - \partial \nu A_\mu$ represents the abelian field-strength tensor and $\epsilon^{\mu \nu \lambda \rho \sigma}$ is the five-dimensional Levi-Civita tensor density with $\epsilon^{01234} = -1$.\\

The five-dimensional G\"{o}del universe is a solution to the equations \eqref{eq:EMCS} with the line element

\begin{align}
\begin{aligned}
\label{eq:linelement}
ds^2 = &- \Big(dt +  j r^2 \left(d\phi + \cos \theta \, d\psi \right)  \Big)^2 + dr^2\\ &+ \frac{r^2}{4} \left(d\theta^2 + d\phi^2 + d\psi^2 + 2 \cos \theta \, d\phi d\psi \right)
\end{aligned}
\end{align}

and the one-form gauge field

\begin{align}
\label{eq:gaugefield}
A_\mu d x^\mu = \frac{\sqrt{3}}{2} jr^2 \left(d\phi + \cos \theta \, d\psi \right),
\end{align}

with $t \in [0,\infty)$, $r\in [0,\infty)$ and Euler angles $\theta \in [0, \pi]$, $\phi \in [0, 2\pi]$ and $\psi \in [0, 4\pi]$. In this metric, the
parameter $j$ defines the scale of the G\"{o}del background and is responsible for the rotation of the universe. For $j=0$ the five-dimensional Minkowski spacetime is recovered. Accordingly, the Kretschmann scalar

\begin{align}
R^{\mu \nu \rho \sigma} R_{\mu \nu \rho \sigma} = 2176 j^4
\end{align}

vanishes for $j=0$. The fact that the Kretschmann scalar is constant reflects the homogeneity of the G\"{o}del spacetime. Calculating the energy-momentum tensor for the gauge field of our solution

\begin{align}
T^{\mu \nu} = 12 j^2 u^\mu u^\nu,
\end{align}

where $u$ is the unit vector in time direction with contravariant components $u^\mu = (1,0,0,0,0)$, one finds that it has vanishing pressure and constant energy density proportional to $j^2$, i.e., the electromagnetic field has the same energy-momentum as pressureless dust. Obviously, the sign of the $g_{\phi \phi}$ component changes for $r > \frac{1}{2j}$, yielding closed timelike curves parameterized by $\phi$ keeping all other coordinates fixed. Note that, since the G\"{o}del spacetime is homogeneous, there is a closed timelike curve through every point in this spacetime.

\subsection{Hamilton-Jacobi equation}

The Hamilton-Jacobi equation for the action $S$, describing a test particle which is coupled to the gauge field \eqref{eq:gaugefield} by a charge $q$, is given by \cite{Misner:1974qy}

\begin{align}
\label{eq:HamJac}
- \frac{\partial S}{\partial \lambda} = \frac{1}{2} g^{\mu \nu} \left(\frac{\partial S}{\partial x^\mu} - q A_\mu \right) \left(\frac{\partial S}{\partial x^\nu} - q A_\nu \right).
\end{align}

Therefore, we need the nonvanishing contravariant metric elements

\begin{align}
\begin{aligned}
g^{tt} &= 4 j^2 r^2 - 1, \quad &g^{rr}&= 1, \quad &g^{\theta \theta} &= \frac{4}{r^2},\\
g^{\phi \phi} &=  \frac{4}{r^2 \sin^2 \theta}, \quad &g^{\psi \psi} &= \frac{4}{r^2 \sin^2 \theta}, \quad &g^{t\phi}&= -4j,\\
g^{\phi \psi} &=  -\frac{4 \cos \theta}{r^2 \sin^2 \theta}.
\end{aligned}
\end{align}

Since the metric has three commuting Killing vectors $\partial_t$, $\partial_\phi$ and $\partial_\psi$, which are related to the conservation of the test particle's energy $E$ and its angular momenta $\Phi$ and $\Psi$, we search for a solution of the form

\begin{align}
\label{eq:Sansatz}
S = \frac{1}{2} \delta \lambda - Et + S_r(r) + S_\theta(\theta) + \Phi \phi + \Psi \psi.
\end{align} 

Here, we introduced $\delta$ as a mass parameter ($\delta = 1$ for massive and $\delta = 0$ for massless test particles), $\lambda$ as the affine parameter along the geodesic and $S_r(r)$, $S_\theta(\theta)$ as being functions depending only on $r$ and $\theta$, respectively. Inserting this ansatz into eq. \eqref{eq:HamJac} yields

\begin{align}
\begin{aligned}
&-\delta r^2 -\left(\frac{\partial S_r}{\partial r} \right)^2 r^2 - 8 E \Phi jr^2 - \left(4 j^2 r^4 - r^2 \right) E^2\\ 
& - 3 j^2 q^2 r^4 + 4 \sqrt{3} j q r^2 \left(Ejr^2 + \Phi \right) = 4\left(\frac{\partial S_\theta}{\partial \theta} \right)^2\\ 
&+ \frac{4}{\sin^2 \theta} \left(\Phi^2 + \Psi^2 - 2 \cos \theta \, \Phi \Psi \right).
\end{aligned} 
\end{align} 

The Hamilton-Jacobi equation is separated into an $r$-dependent left-hand and a $\theta$-dependent right-hand side. Thus, we can set both sides equal to a separation constant $K$ resulting in two equations

\begin{align}
\begin{aligned}
r^2 \left(\frac{\partial S_r}{\partial r} \right)^2 &= - K -\delta r^2 - 8 E \Phi j r^2 - \left(4 j^2 r^4 - r^2 \right) E^2 \\
 &\quad\,- 3 j^2 q^2 r^4 + 4 \sqrt{3} j q r^2 \left(Ejr^2 + \Phi \right) =: R
\end{aligned} 
\end{align} 

and

\begin{align}
\left(\frac{\partial S_\theta}{\partial \theta} \right)^2 &= \frac{K}{4} - \frac{1}{\sin^2 \theta} \left(\Phi^2 + \Psi^2 - 2 \cos \theta \, \Phi \Psi \right) =: \Theta.
\label{eq:Thetadef}
\end{align} 

The right-hand side functions $R$ and $\Theta$ have been introduced for brevity. The action \eqref{eq:Sansatz} now takes the form

\begin{align}
\label{eq:Sansatz}
S = \frac{1}{2} \delta \lambda - Et +\epsilon_r \int^r\frac{\sqrt{R}}{r} \, dr +\epsilon_\theta \int^\theta \sqrt{\Theta} \, d\theta + \Phi \phi + \Psi \psi,
\end{align} 

where $\epsilon_r$ and $\epsilon_\theta$ refer to the independent signs of the square roots. Differentiating this action with respect to the constants of motion $K$, $\delta$, $\Phi$, $\Psi$ and $E$ and setting the resulting constants equal to zero yields the geodesic equations

\begin{align}
\label{eq:geor}
\left(\frac{dr}{d\tau} \right)^2 &= R r^2,\\
\label{eq:geotheta}
\left(\frac{d\theta}{d\tau} \right)^2 &= 16 \Theta,\\
\left(\frac{d\phi}{d\tau} \right) &= 4jr^2 \left(E - \frac{\sqrt{3}}{2} q\right) + 4 \frac{\Phi - \Psi \cos \theta}{\sin^2 \theta},\\
\label{eq:geopsi}
\left(\frac{d\psi}{d\tau} \right) &= 4 \frac{\Psi - \Phi \cos \theta}{\sin^2 \theta},\\
\label{eq:geot}
\left(\frac{dt}{d\tau} \right) &= \left(2 \sqrt{3} q - 4 E j^2 \right) r^4 + \left(E - 4 \Phi j \right) r^2 ,
\end{align} 

where we introduced a new parameter $\tau$ along the geodesic by \cite{Mino:2003yg}

\begin{align}
d \tau = \frac{d\lambda}{r^2}.
\end{align} 

Obviously, the $\theta$ and $\psi$ motions are not affected by the test particle's charge $q$ and the rotation parameter $j$. As well as the $r$ and $t$ motions are not affected by the test particle's angular momentum $\Psi$.
    
\section{Discussion of the motion}
\label{sec:disc_motion}

The obtained geodesic equations \eqref{eq:geor} - \eqref{eq:geot} allow us to investigate the motion of test particles qualitatively by studying their right-hand sides. 

\subsection{$\theta$ motion}

The $\theta$ motion is described by Eq. \eqref{eq:geotheta}. Obviously, the subspace $\theta=0$ or $\theta=\pi$, respectively, can only be reached if $\Phi=\pm \Psi$. Other constant $\theta$ motions are determined by 

\begin{align}
	\Theta(\theta_0) = 0 \qquad \text{and} \qquad \left.\frac{d\Theta}{d\theta}\right\vert_{\theta=\theta_0} = 0.
	\label{eq:constanttheta}
\end{align} 

In order to simplify the calculations, we transform Eq.\eqref{eq:geotheta} by substituting

\begin{align}
	\xi=\cos \theta, \qquad \xi \in [-1,1]
	\label{eq:thetasubs}
\end{align} 

yielding a polynomial of the form

\begin{align}
	\left(\frac{d\xi}{d\tau} \right)^2 = a_2 \xi^2 + a_1 \xi + a_0 =:\Xi,
	\label{eq:xiequation}
\end{align} 

where

\begin{align}
	\begin{aligned}
		a_2 &= - 4K,\\
		a_1 &= 32\Phi \Psi,\\
		a_0 &= 4 K - 16 \Phi^2 - 16 \Psi^2,
	\end{aligned} 
\end{align} 

with the discriminant

\begin{align}
	D_\xi = a_1^2 - 4 a_2 a_0 = 64 (K-4\Phi^2) (K-4\Psi^2).
\end{align}

Therefore, Eqs. \eqref{eq:constanttheta} are equivalent to

\begin{align}
	\Xi(\xi_0) = 0 \qquad \text{and} \qquad \left.\frac{d\Xi}{d\xi}\right\vert_{\xi=\xi_0} = 0.
	\label{eq:constantxi}
\end{align}

For $K \neq 0$ these equations can be fulfilled by

\begin{align}
	D_\xi = 64 (K-4\Phi^2) (K-4\Psi^2) \overset{!}{=} 0,
\end{align} 

such that we obtain a constant $\theta$ motion if $K=4\Phi^2$ or $K=4\Psi^2$. For $K=0$ we use that Eq. \eqref{eq:Thetadef} requires

\begin{align}
	\frac{1}{\sin^2 \theta} \left(\Phi^2 + \Psi^2 - 2 \cos \theta \, \Phi \Psi \right) \le 0.
\end{align} 

Since this term is non-negative, it must vanish, which is fulfilled iff $\theta=0$ and $\Phi=\Psi$ or $\theta=\pi$ and $\Phi = - \Psi$. A non-constant $\theta$ motion is bounded by the zeros of $\Theta$ or $\Xi$, respectively. For the motion to be physical we require that the zeros $\xi_0^{1,2}$ must be real, i.e. $D_\xi >0 $ and $\xi \in [-1,1]$ due to the transformation \eqref{eq:thetasubs}. Furthermore, $\Xi$ must be positive between these zeros in order to yield a physical motion with some real $\xi(\tau)$ and thus $\theta(\tau)$. We will investigate the behavior of the zeros of $\Theta$ by its discriminant

For $K> 0$, the roots of $\Xi$ determine the turning points of a non-constant $\theta$ motion and, therefore, need to be real. Consequently, the discriminant must be non-negative, which is true for the two cases

\begin{align}
	K \ge 4\Phi^2 \quad  \cup \quad  K \ge 4\Psi^2
	\label{eq:Krestriction}
\end{align} 

or

\begin{align}
	0< K < 4\Phi^2 \quad  \cup \quad 0 < K < 4\Psi^2.
\end{align} 

Fig. \ref{fig:thetadiscrim} illustrates the discriminant as a function of $\Phi$ and $\Psi$ in case of $K=2$:

\begin{figure}[H]
	\centering
	\includegraphics[width=0.4\textwidth]{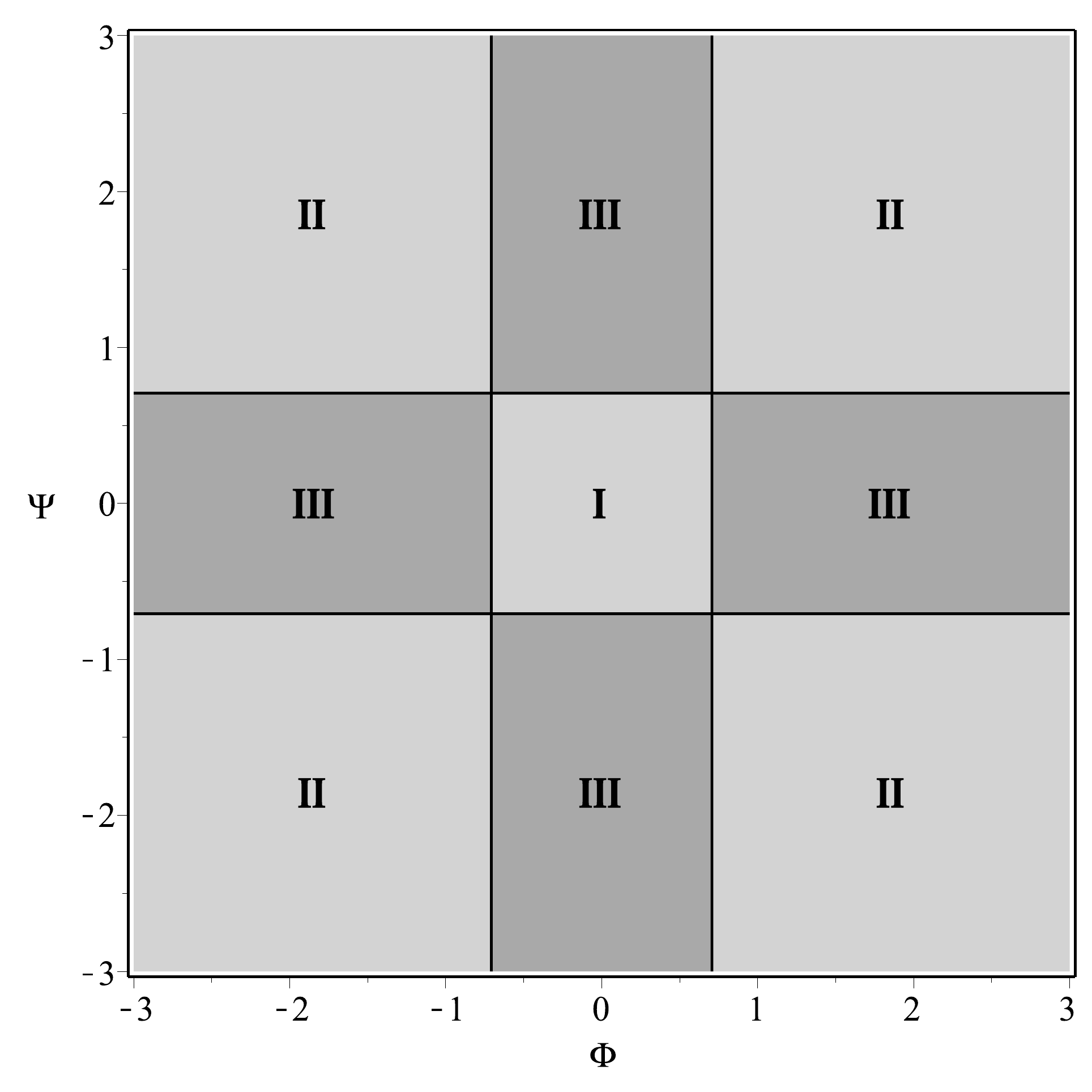}
	\caption{Parametric $\Phi$-$\Psi$-plot of the $\Xi$ discriminant for $K = 2$.}
	\label{fig:thetadiscrim}
\end{figure}

The indicated regions are related to the zeros of $\Xi$ in the way shown in Tab. \ref{tab:thetaregions}, where we excluded the special cases $\Phi = \pm \Psi$. The number of physical turning points is confined by $\xi_0^{1,2} \in [-1,1]$ due to Eq. \eqref{eq:thetasubs}:

\begin{table}[H]
	\centering
	\begin{tabular}{|c|c|c|}
		\hline
		Region & Number of real zeros & Number of zeros $\in (-1,1)$\\
		\hline
		\hline
		I   & $2 \in \mathbb{R}$  & 2 \\
		\hline
		II  & $2 \in \mathbb{R}$  &  0 \\
		\hline
		III & $0 \in \mathbb{R}$ &  0\\
		\hline
	\end{tabular}
	\caption{Zeros of $\Xi$ for different regions of the $E$-$\Phi$-plots.}
	\label{tab:thetaregions}
\end{table}

\begin{figure}[H]
	\centering
	\includegraphics[width=0.4\textwidth]{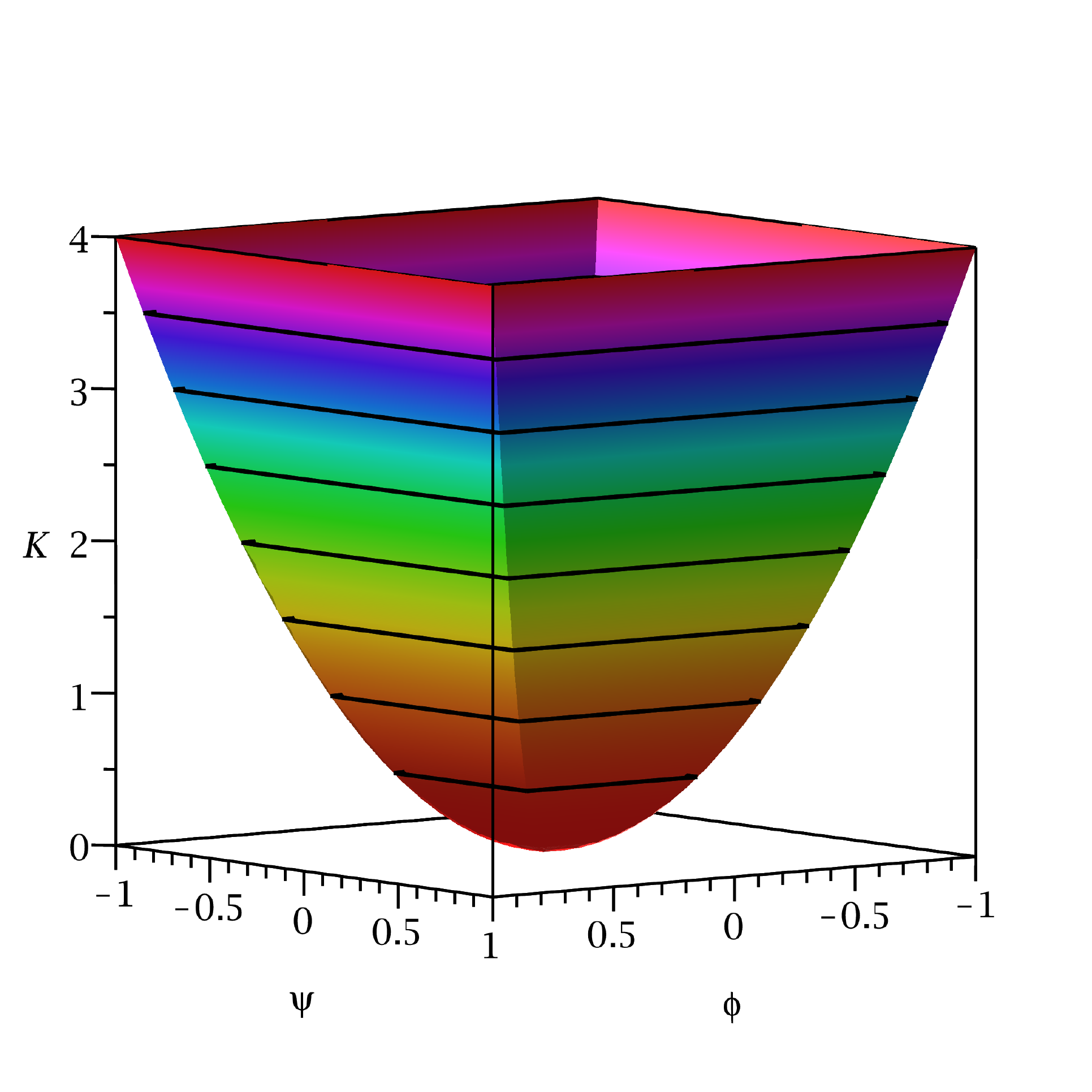}
	\caption{Parametric plot of the $\Xi$ discriminant.}
	\label{fig:thetadiscrim3d}
\end{figure}

Consequently, only the values of angular momenta in region I are related to a physical $\theta$ motion, which are given by Eq. \eqref{eq:Krestriction}. We can visualize the boundary of this region for different values of $K$ in a three-dimensional plot as shown in Fig. \ref{fig:thetadiscrim3d}.

Parameter values for $K$, $\Phi$ and $\Psi$ inside this boundary are related to a non-constant $\theta$ motion and those on the boundary are related to a constant $\theta$ motion. Other values do not yield a physical motion.

\subsection{r motion}

The radial motion is determined by Eq. \eqref{eq:geor}

\begin{align}
\left(\frac{dr}{d\tau} \right)^2 &= R r^2.
\end{align}

Again, we can conclude that $R$ must be positive in order to yield a physical motion, where the real zeros of the right-hand side denote the radial turning points. Obviously, $r=0$ will always be a double zero, but for small values of $r$ the only relevant coefficient is given by $-K$. Since we have already proven $K\ge 0$, $r=0$ may only be reached with positive $R$ iff $K=0$. In case of $K\neq 0$, there are either two or zero positive roots of $R$, due to Descartes' rule of signs.  Since $r\in [0,\infty)$, only the positive zeros are physically valid. In the case of two radial turning points $r_1, r_2$ we have bound orbits (BO) with range $r \in [r_1, r_2]$ and $0 < r_{1} < r_{2}$. In the special case of 

\begin{align}
E = \frac{\sqrt{3}}{2} q
\end{align}

the leading coefficient of $R$ vanishes and therefore $R$ reduces to a quadratic polynomial. In this case, $R$ either has one or zero positive roots. In the case of a single radial turning point $r_1$ we have escape orbits (EO) with range $r$ $\in$ [$r_{1}$,$\infty$). \\

A very instructive way of investigating the radial motion is given by the effective potential. Therefore we rewrite the radial equation as follows

\begin{align}
\left(\frac{dr}{d\tau} \right)^2 = \gamma_2 E^2 + \gamma_1 E + \gamma_0,
\end{align}

where

\begin{align}
\begin{aligned}
\gamma_2 &= r^4 - 4 j^2 r^6,\\
\gamma_1 &= 4\sqrt{3} j^2 q r^6 - 8\Phi j r^4,\\
\gamma_0 &= 4\sqrt{3}\Phi j q r^4 - 3 j^2 q^2 r^6 - \delta r^4 - K r^2.
\end{aligned}
\end{align}

The zeros of this quadratic polynomial are given by

\begin{align}
V_{\rm eff}^\pm := \frac{-\gamma_1 \pm \sqrt{\gamma_1^2 - 4\gamma_2 \gamma_0}}{2 \gamma_2}
\end{align}

and they define the two branches of an effective potential $V_{\rm eff}$ for the test particle's energy

\begin{align}
\dot r^2 = \gamma_2 \left(E-V_{\rm eff}^+\right) \left(E-V_{\rm eff}^- \right).
\end{align}

The radial turning points are now given by $E = V_{\rm eff}^\pm$, so that we may easily visualize and characterize the possible orbit types as presented in Fig. \ref{fig:goedelpotentials}:

\begin{figure*}[tp]
\centering
\begin{minipage}[htbp]{0.33\textwidth}
   \includegraphics[width=\textwidth]{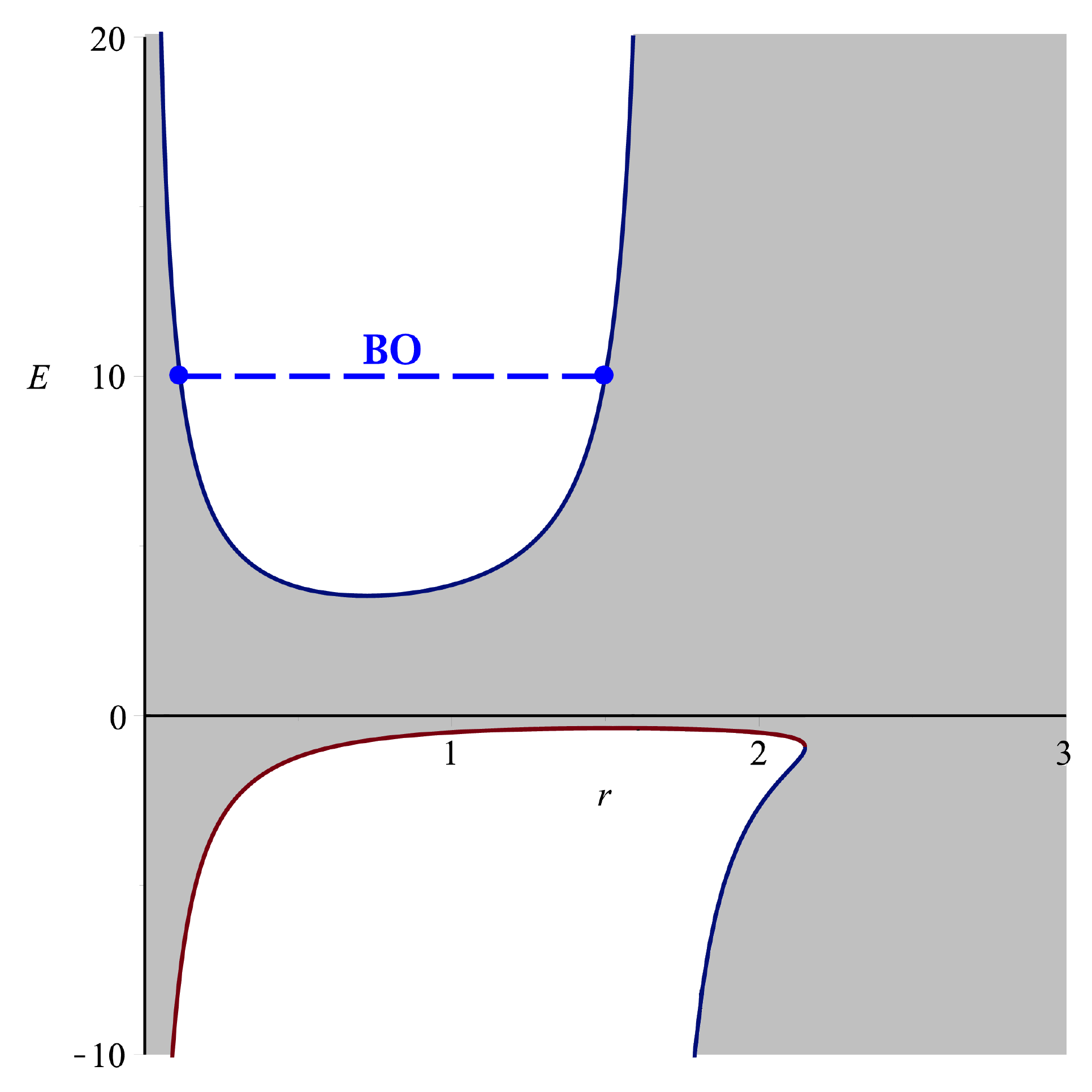}
\subcaption{\footnotesize $\delta = 1,j = 0.3, q = 0.4, K = 1, \Phi = 1$.}
   \end{minipage}\hfill
   \begin{minipage}[htbp]{0.33\textwidth}
   \includegraphics[width=\textwidth]{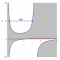}
\subcaption{\footnotesize $\delta = 0,j = 0.3, q = 0, K = 1, \Phi = 1$.}
   \end{minipage}\hfill
   \begin{minipage}[htbp]{0.33\textwidth}
   \includegraphics[width=\textwidth]{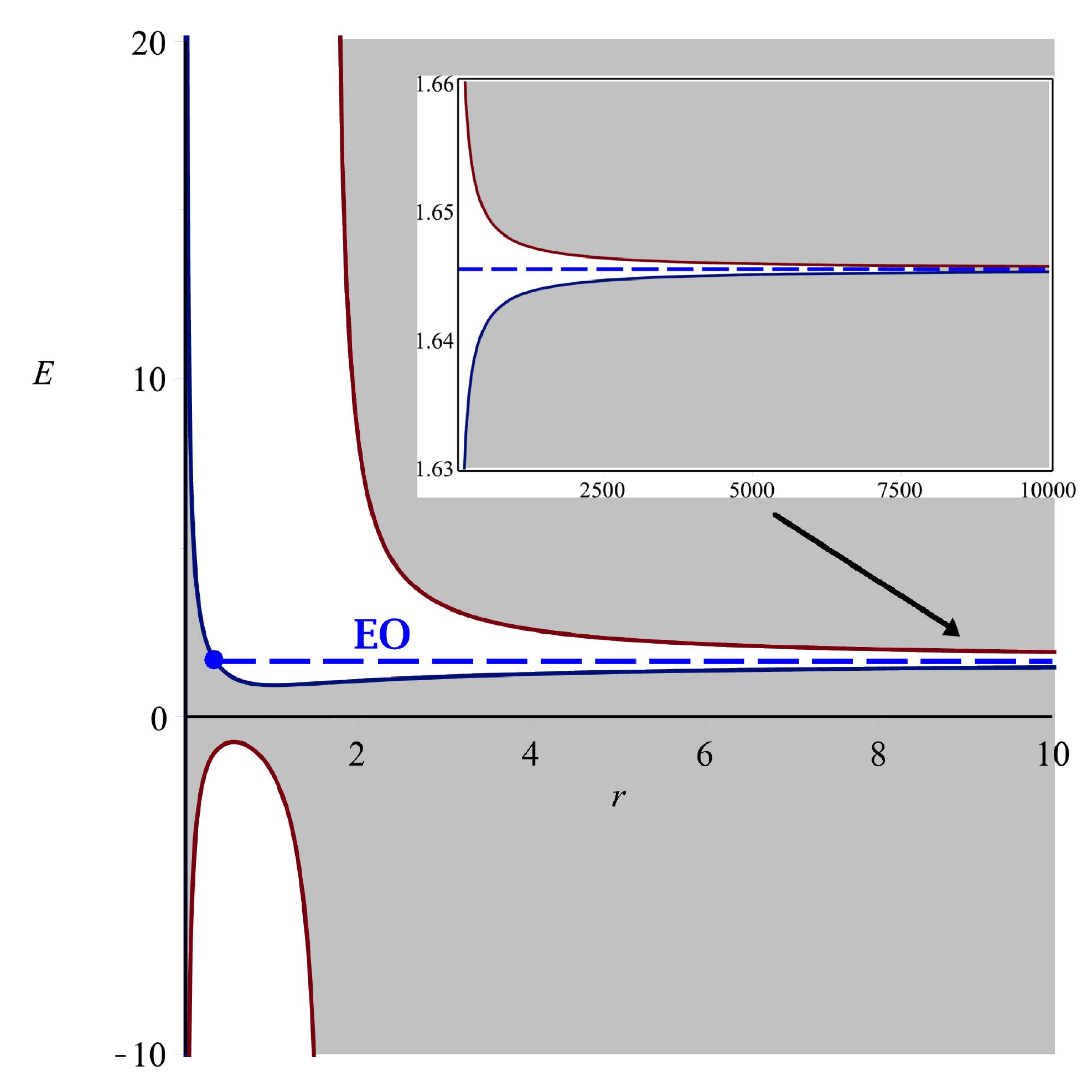}
\subcaption{\footnotesize $\delta = 1,j = 0.3, q = 1.9, K = 0.2, \Phi = 0.3$.}
   \end{minipage}

\begin{minipage}[htbp]{0.33\textwidth}
   \includegraphics[width=\textwidth]{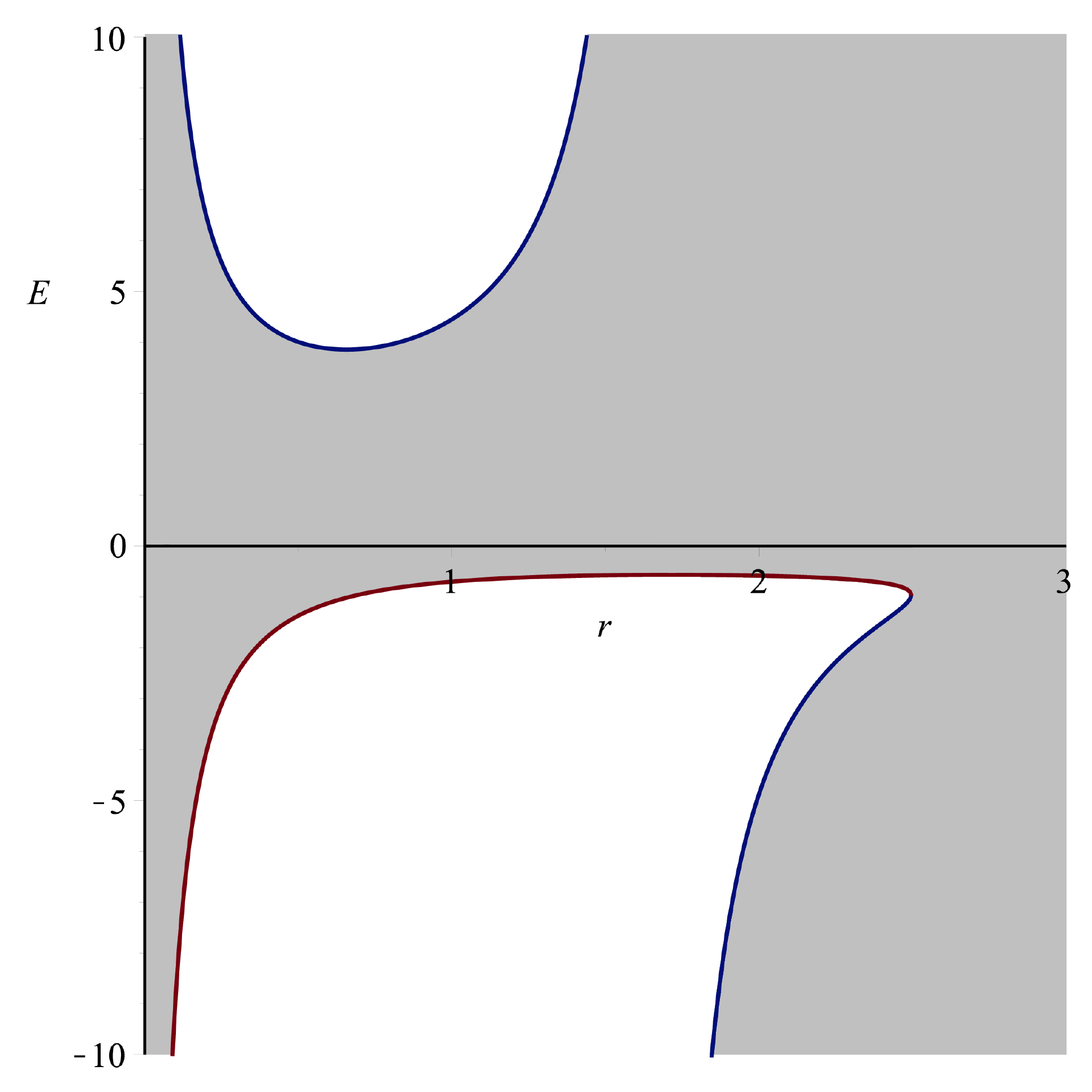}
   \subcaption{\footnotesize $\delta = 1,j = 0.3, q =- 0.4, K = 1, \Phi = 1$.}
   \end{minipage}\hfill
   \begin{minipage}[htbp]{0.33\textwidth}
   \includegraphics[width=\textwidth]{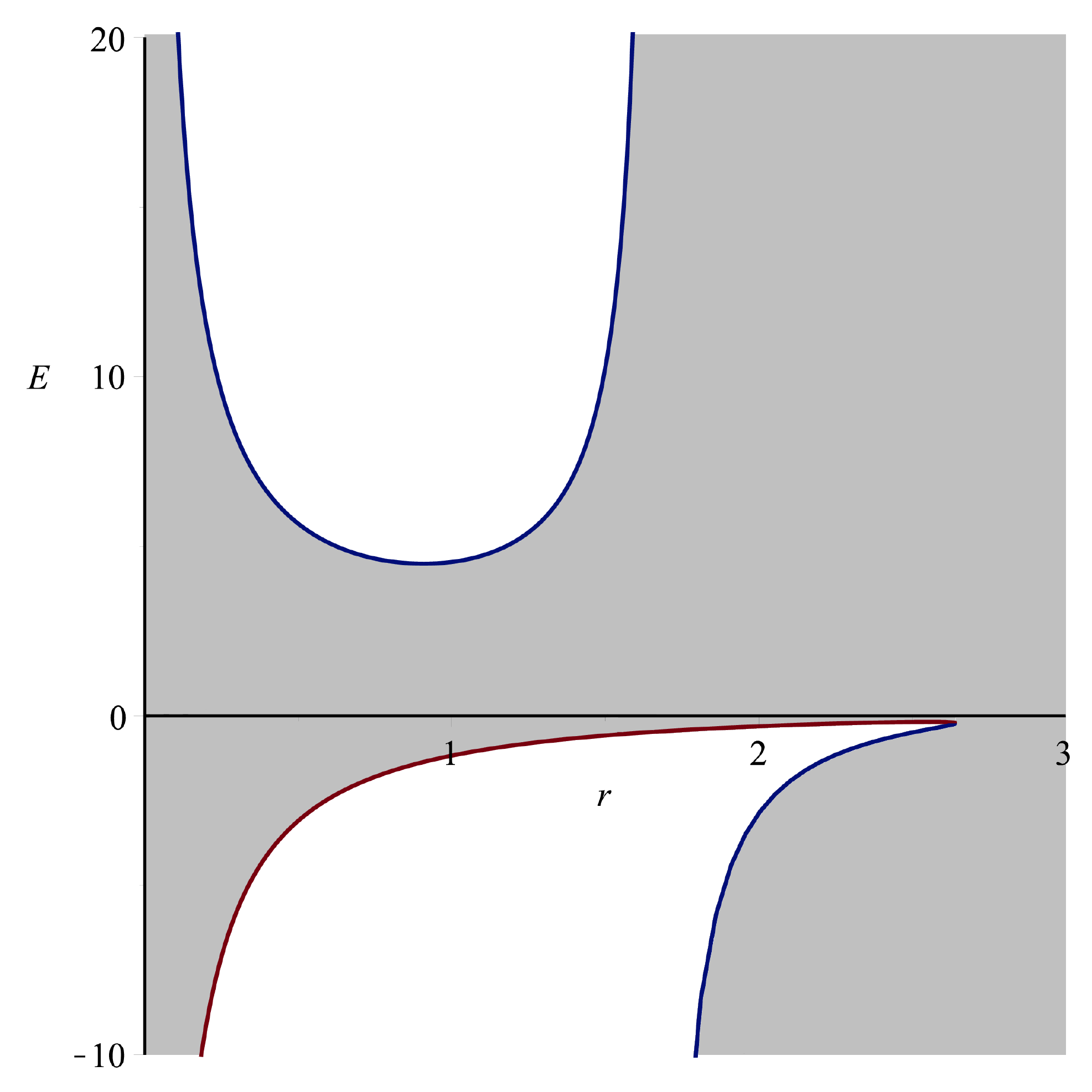}
   \subcaption{\footnotesize $\delta = 1,j = 0.3, q = 0, K = 1, \Phi = 1$.}
   \end{minipage}\hfill
   \begin{minipage}[htbp]{0.33\textwidth}
   \includegraphics[width=\textwidth]{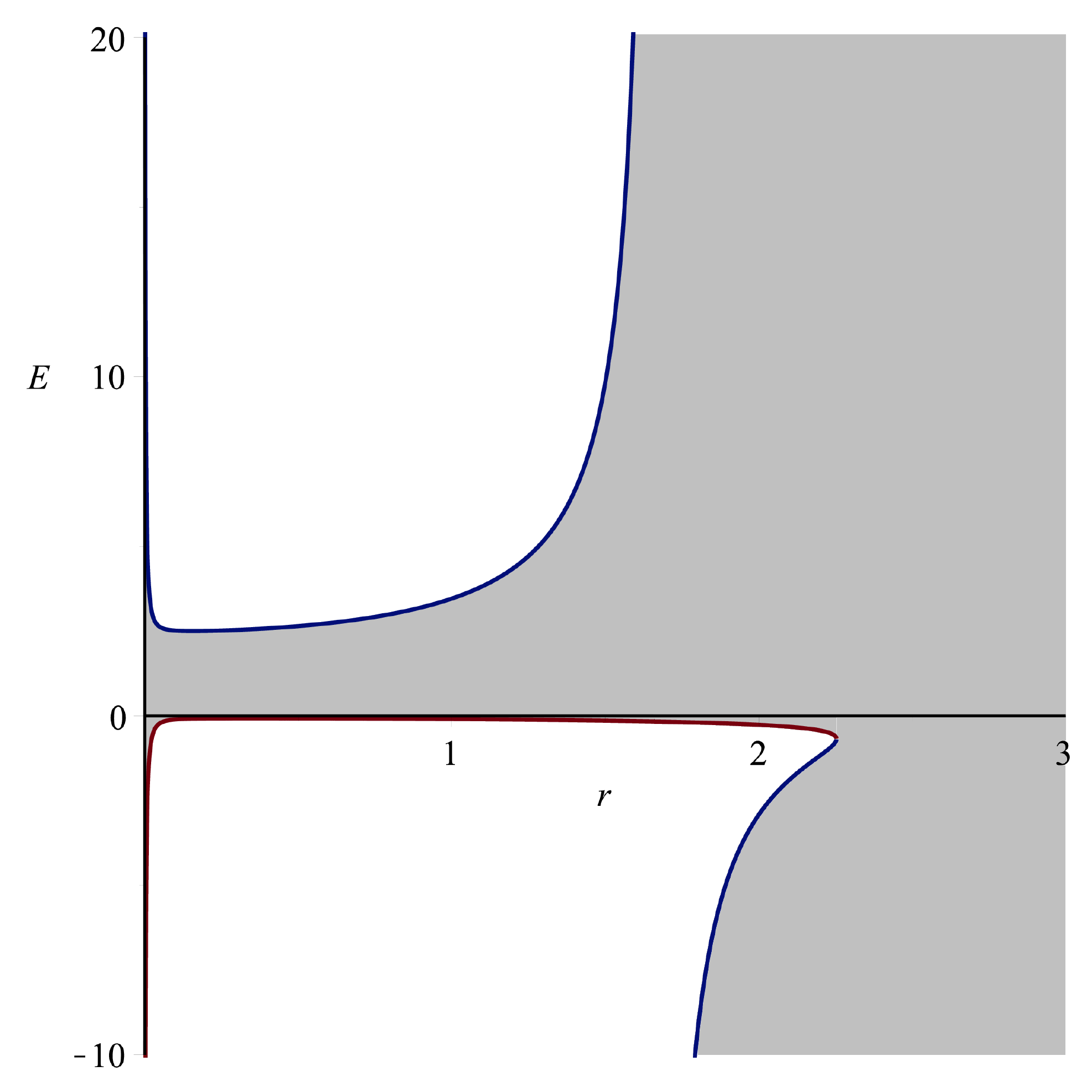}
   \subcaption{\footnotesize $\delta = 1,j = 0.3, q = 0.4, K = 0.001, \Phi = 1$.}
   \end{minipage}
   \caption{Effective potentials (blue, red) for the radial test particle and light motion in the five-dimensional G\"{o}del spacetime. The blue dashed lines denote the energy of the related  orbit and the blue dots mark the zeros of the radial polynomial $R$, which are the radial turning points of the orbits. In the grey area no motion is possible since $R<0$. The only possible orbit types are bound and escape orbits.}
\label{fig:goedelpotentials}
\end{figure*}

In Fig. \ref{fig:orbittypes} all possible types of orbits are summarized:

\begin{figure}[H]
\centering
\includegraphics[width=0.49\textwidth]{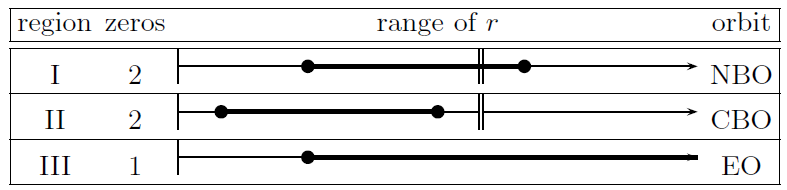}
\caption{Representation of the orbit types for massive ($\delta =1$) and massless ($\delta =0$) test particles together with the number of zeros in the respectively region. The noncausal bound orbits (NBO) cross the modified G\"{o}del sphere which is given as a vertical bar in the plots. The EOs are excluded for uncharged lightlike motion.}
\label{fig:orbittypes}
\end{figure}

\subsection{t motion}
\label{sec:t}

The $t$ equation is given by Eq. \eqref{eq:geot}. Due to causality, the right-hand side must be positive. Therefore, we calculate the zeros 

\begin{align}
r_{1,2} = 0, \qquad r_{3,4} =\pm \frac{1}{2} \sqrt{\frac{E-4\Phi j}{E j^2 - \frac{\sqrt{3}}{2} q}}
\end{align} 

and conclude that either 

\begin{align}
E-4\Phi j >0 \quad  \cup \quad  E j^2 - \frac{\sqrt{3}}{2} q >0
\end{align} 

or

\begin{align}
E-4\Phi j <0 \quad  \cup \quad  E j^2 - \frac{\sqrt{3}}{2} q <0
\end{align} 

must be fulfilled. The latter case results in a positive leading coefficient for the right-hand side of the $t$ equation yielding $\dot t<0$  between $r=0$ and $r=r_3$. In order to obtain a physical motion, we will restrict the parameters to the first case. The first case leads to a modified G\"{o}del-radius $r_3$, which contains causal bound orbits (CBO). The bound orbits which pass this radius are noncausal (NBO) and can lead to the existence of CTGs. A detailled discussion is given in \ref{sec:orb}. For $q=\Phi =0$ this modified G\"{o}del radius becomes exactly the classical one.

\section{Analytical solutions}

In this section, we will solve the geodesic equations \eqref{eq:geor} - \eqref{eq:geot}.

\subsection{$\theta$ equation}

In order to solve the $\theta$ equation we will use the substitution $\xi=\cos \theta$ again, which led to Eq. \eqref{eq:xiequation}

\begin{align}
\dot \xi^2 = a_2 \xi^2 + a_1 \xi + a_0.
\end{align} 

Separation of variables leads to an integral of the form

\begin{align}
\tau - \tau_{\rm in} = \int_{\xi_{\rm in}}^\xi \frac{d\xi}{\sqrt{a_2 \xi^2 + a_1 \xi + a_0}},
\label{eq:xiintegral}
\end{align} 

where $\xi_{\rm in} = \xi(\tau_{\rm in})$. For a physical, non-constant $\theta$ motion we already showed that the  discriminant $D_\xi$ should be positive and the leading coefficient $a_2$ should be negative. In this case the integral \eqref{eq:xiintegral} yields \cite{Gradshteyn:2007}

\begin{align}
\tau - \tau_{\rm in}^\theta = -\frac{1}{\sqrt{-a_2}} \arcsin \left(\frac{2a_2 \xi + a_1}{\sqrt{D_\xi}} \right), 
\end{align} 

where

\begin{align}
\tau_{\rm in}^\theta = \tau_{\rm in} + \frac{1}{\sqrt{-a_2}} \arcsin \left(\frac{2a_2 \xi_{\rm in} + a_1}{\sqrt{D_\xi}} \right).
\end{align} 

Solving this equation for $\xi$ and resubstitute $\xi=\cos \theta$ gives the final solution

\begin{align}
\theta(\tau) =\arccos \left(-\frac{\sqrt{D_\xi}}{2 a_2} \sin \Big(\sqrt{-a_2} \left(\tau - \tau_{\rm in}^\theta \right)\Big) - \frac{a_1}{2 a_2}\right).
\end{align} 

\subsection{$r$ equation}

In order to solve the $r$ equation \eqref{eq:geor} analytically we perform a substitution via

\begin{align}
x = \frac{1}{r^2}
\label{eq:radialsub}
\end{align}

yielding

\begin{align}
\left(\frac{dx}{d\tau} \right)^2 &=  b_2 x^2 + b_1 x + b_0 =:\mathcal{X},
\label{eq:xequation}
\end{align}

where 

\begin{align}
\begin{aligned}
b_2 &= -4K,\\
b_1 &= 4\left( E^2 - \delta \right) - 32 E \Phi j + 16 \sqrt{3} \, \Phi jq,\\
b_0 &= - 4 \left(2 E - \sqrt{3} q \right)^2 j^2.
\end{aligned}
\end{align}

Separation of variables leads to same integral as for the $\theta$ equation \eqref{eq:xiintegral} resulting in

\begin{align}
x(\tau) = -\frac{\sqrt{D_x}}{2 b_2} \sin \left(\sqrt{-b_2}\left(\tau - \tau_{\rm in}^x \right) \right)- \frac{b_1}{2 b_2},
\end{align} 

where $D_x$  is the discriminant of $X$ and

\begin{align}
\tau_{\rm in}^r = \tau_{\rm in} - \frac{1}{\sqrt{-b_0}} \arcsin \left(\frac{2b_0 + b_1 x_{\rm in}}{x\sqrt{D_x}} \right).
\end{align} 

Therefore, the $r$ equation is finally solved by 

\begin{align}
\label{eq:rsol}
r(\tau) =\sqrt{\frac{-2b_2}{\sqrt{D_x} \sin \Big(\sqrt{-b_2} \left(\tau - \tau_{\rm in}^\theta \right)\Big) + b_1}}.
\end{align} 

\subsection{$\phi$ equation}

The $\phi$ equation consists of a  $\theta$- and an $r$-dependent part. Separation of variables and substituting \eqref{eq:geor} and \eqref{eq:geotheta} as well as \eqref{eq:xiequation} and \eqref{eq:xequation} yields

\begin{align}
\begin{aligned}
d\phi &= 4jr^2 \left(E - \frac{\sqrt{3}}{2} q\right)d\tau + 4 \frac{\Phi - \Psi \cos \theta}{\sin^2 \theta} d\tau\\
&= 4j \left(E - \frac{\sqrt{3}}{2} q\right) \frac{dx}{x\sqrt{X}}+ 4 \frac{\Phi - \Psi \xi}{1-\xi^2} \frac{d\xi}{\sqrt{\Xi}}\\
&=: d\phi_x + d\phi_\xi.
\end{aligned} 
\end{align} 

The integration of $d\phi_x$ is straightforward and yields \cite{Gradshteyn:2007}

\begin{align}
\begin{aligned}
\phi_x(\tau) &=  \frac{4j}{\sqrt{-b_0}} \left(E - \frac{\sqrt{3}}{2} q\right) \bigg[ \arcsin \left(\frac{2b_0 r^2(\tau)+b_1}{\sqrt{D_x}} \right)\\
& \quad -  \arcsin \left(\frac{2b_0 r_{\rm in}^2+b_1}{\sqrt{D_x}} \right) \bigg] + \phi_{\rm in}^x,
\end{aligned}
\end{align}

where $ r_{\rm in} = r(\tau_{\rm in})$. In order to integrate  $d\phi_\xi$ we need to perform a partial fraction decomposition

\begin{align}
d\phi_\xi = 4 \frac{\Phi - \Psi \xi}{1-\xi^2} \frac{d\xi}{\sqrt{\Xi}} = 2 \frac{\Psi + \Phi}{\xi+1} \frac{d\xi}{\sqrt{\Xi}} + 2 \frac{\Psi - \Phi}{\xi-1} \frac{d\xi}{\sqrt{\Xi}}
\label{eq:phipartialdecomp}
\end{align} 

yielding two integrable parts. Using the substitutions $t^\pm= \pm \frac{1}{\xi\pm 1} > 0$, for the first and the second term, respectively, yields \cite{Gradshteyn:2007}

\begin{align}
	\int_{\xi_{\rm in}}^\xi \frac{\left(\Psi \pm \Phi \right)d\xi}{(\xi\pm 1) \sqrt{\Xi}} =\mp \int_{t^\pm_{\rm in}}^{t^\pm} \frac{\left(\Psi \pm \Phi \right)dt^\pm}{\sqrt{ c_2 t^2 + c_1 t + c_0}},
\end{align} 

where

\begin{align}
	\begin{aligned}
		c_0^\pm &= a_2\\
		c_1^\pm &= a_1 \mp 2a_2\\
		c_2^\pm &= a_0 -\mp a_1 + a_2.
	\end{aligned} 
\end{align} 

Since the value of the discriminant remains unchanged, we can apply the same solution method as used for the $\theta$ and $r$ equation, resulting in

\begin{align}
\begin{aligned}
\phi(\tau) =&-\frac{2(\Psi+\Phi)}{\sqrt{-c_2^+}} \arcsin \left(\frac{2 c_2^+ t^+(\tau) + c_1^+}{\sqrt{D_\xi}} \right)\\
&+\frac{2(\Psi+\Phi)}{\sqrt{-c_2^+}} \arcsin \left(\frac{2 c_2^+ t^+_{\rm in} + c_1^+}{\sqrt{D_\xi}} \right)\\
&+\frac{2(\Psi-\Phi)}{\sqrt{-c_2^-}} \arcsin \left(\frac{2 c_2^- t^-(\tau) + c_1^-}{\sqrt{D_\xi}} \right)\\
&-\frac{2(\Psi-\Phi)}{\sqrt{-c_2^-}} \arcsin \left(\frac{2 c_2^- t^-_{\rm in} + c_1^-}{\sqrt{D_\xi}} \right) + \phi_{\rm in}.
\end{aligned} 
\end{align}

\subsection{$\psi$ equation}

The $\psi$ equation, which depends solely on the $\theta$ equation, is given by Eq. \eqref{eq:geopsi}. Separation of variables and substitution of \eqref{eq:xiequation} leads to

\begin{align}
d\psi = 4 \frac{\Psi - \Phi \cos \theta}{\sin^2 \theta} d\tau =  4 \frac{\Psi - \Phi \xi}{1-\xi^2} \frac{d\xi}{\sqrt{\Xi}}.
\end{align} 

This differential is of the same form as \eqref{eq:phipartialdecomp} when $\Phi$ and $\Psi$ are exchanged. Consequently, we can give the solution directly by

\begin{align}
	\begin{aligned}
		\psi(\tau) =&-\frac{2(\Phi+\Psi)}{\sqrt{-c_2^+}} \arcsin \left(\frac{2 c_2^+ t^+(\tau) + c_1^+}{\sqrt{D_\xi}} \right)\\
		&+\frac{2(\Phi+\Psi)}{\sqrt{-c_2^+}} \arcsin \left(\frac{2 c_2^+ t^+_{\rm in} + c_1^+}{\sqrt{D_\xi}} \right)\\
		&+\frac{2(\Phi-\Psi)}{\sqrt{-c_2^-}} \arcsin \left(\frac{2 c_2^- t^-(\tau) + c_1^-}{\sqrt{D_\xi}} \right)\\
		&-\frac{2(\Phi-\Psi)}{\sqrt{-c_2^-}} \arcsin \left(\frac{2 c_2^- t^-_{\rm in} + c_1^-}{\sqrt{D_\xi}} \right) + \psi_{\rm in}.
	\end{aligned} 
\end{align}

\section{Orbits}
\label{sec:orb}

The coordinates $(r,\theta, \phi, \psi)$ are related to cartesian coordinates in $\mathbb{R}^4$ via \cite{Gauntlett:2002nw}

\begin{align}
\begin{aligned}
X + i Y &= r \cos \frac{\theta}{2} \, e^{\frac{i}{2} \left(\psi + \phi \right)}\\
Z + i W &= r \sin \frac{\theta}{2} \, e^{\frac{i}{2} \left(\psi - \phi \right)}.
\end{aligned}
\end{align}

\begin{figure}[h]
\begin{subfigure}[H]{.5\linewidth}
	\centering
	\includegraphics[width=0.99\textwidth]{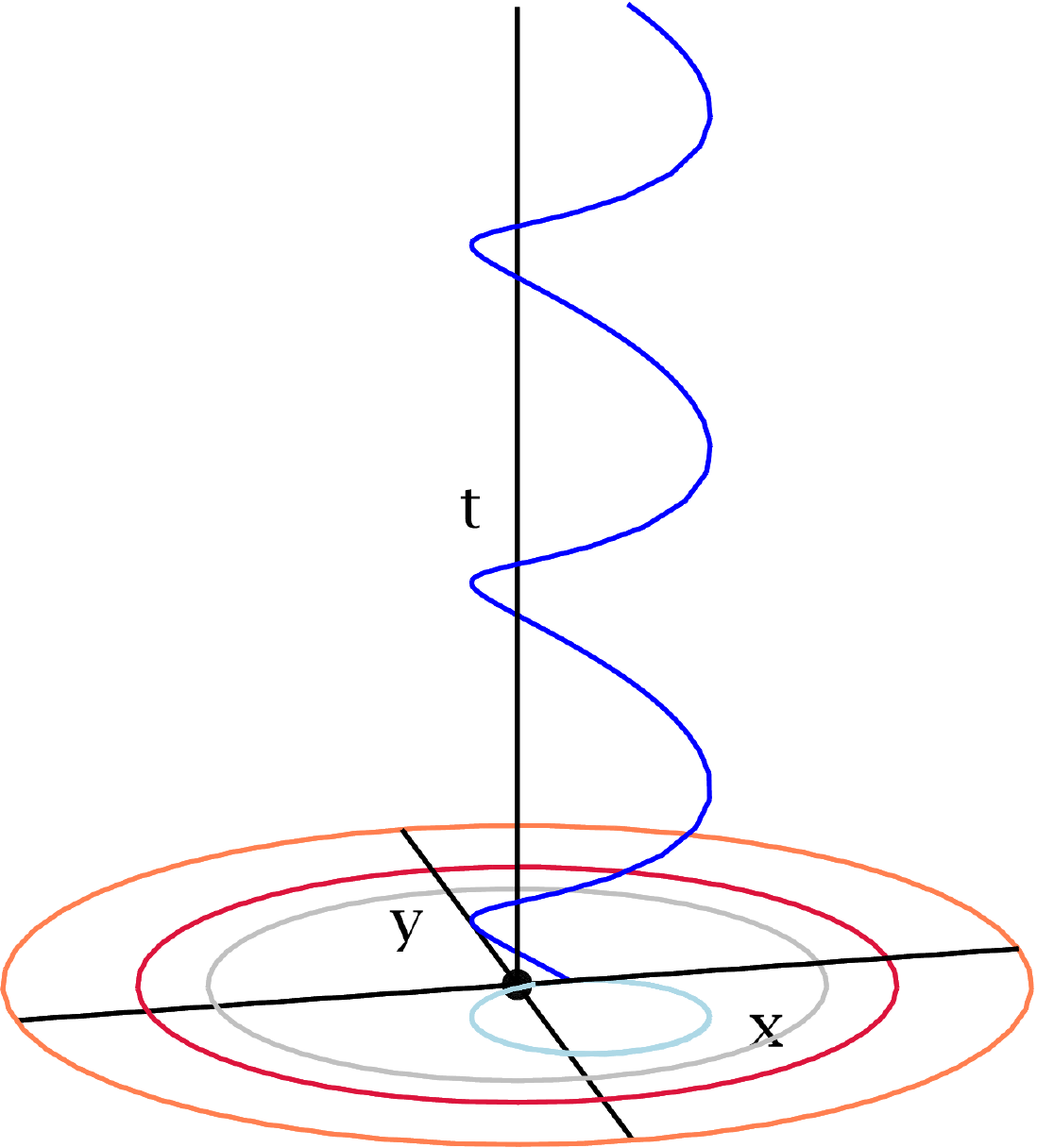}
	\caption{\centering \footnotesize $X$-$Y$-$t$-projection.}
\end{subfigure}%
\begin{subfigure}[H]{0.49\linewidth}
    \centering
	\includegraphics[width=0.99\textwidth]{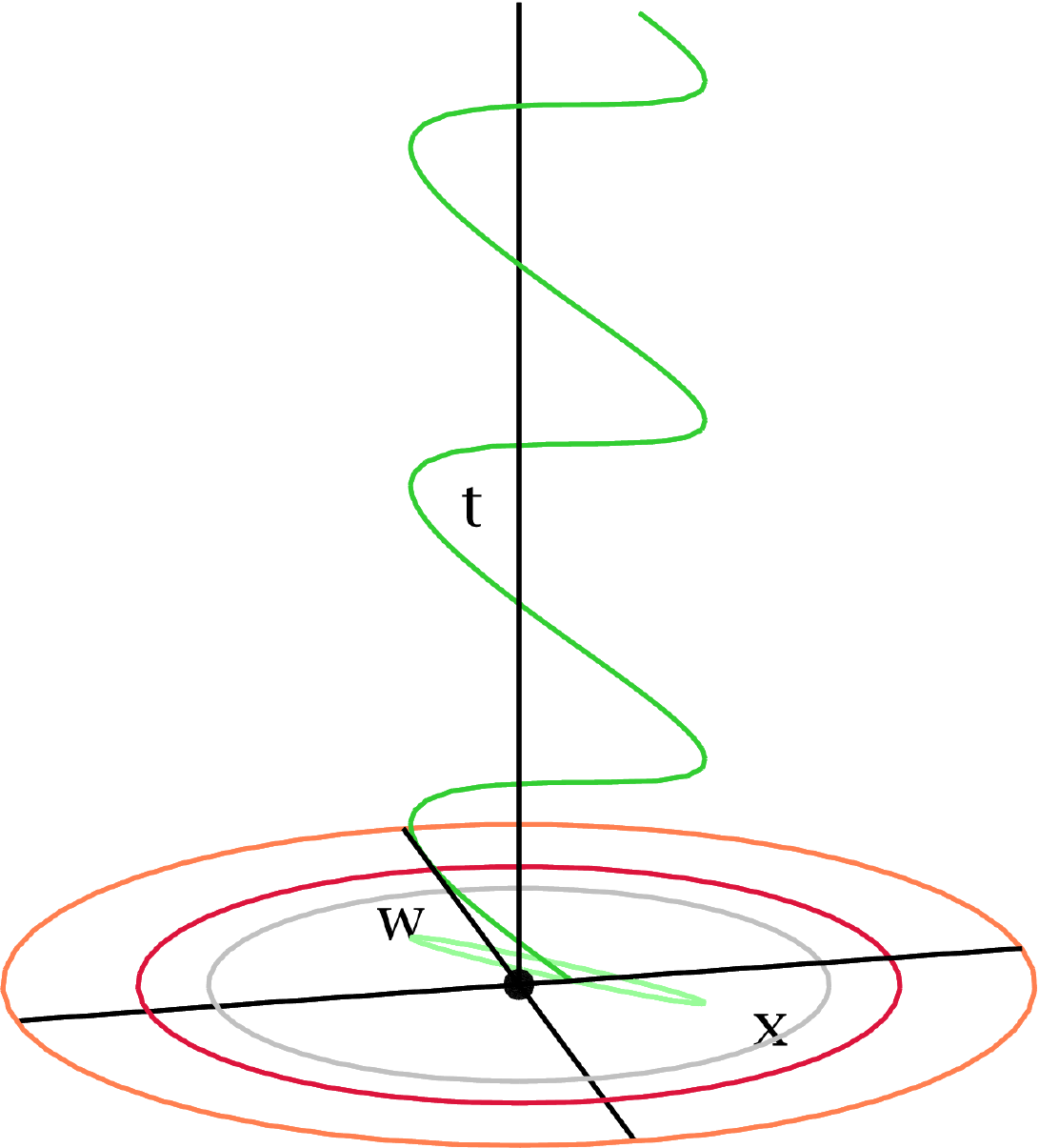}
	\caption{\centering \footnotesize $X$-$W$-$t$-projection.}
\end{subfigure}\\[20pt]	

\begin{subfigure}[H]{0.49\linewidth}
	\centering
	\includegraphics[width=0.99\textwidth]{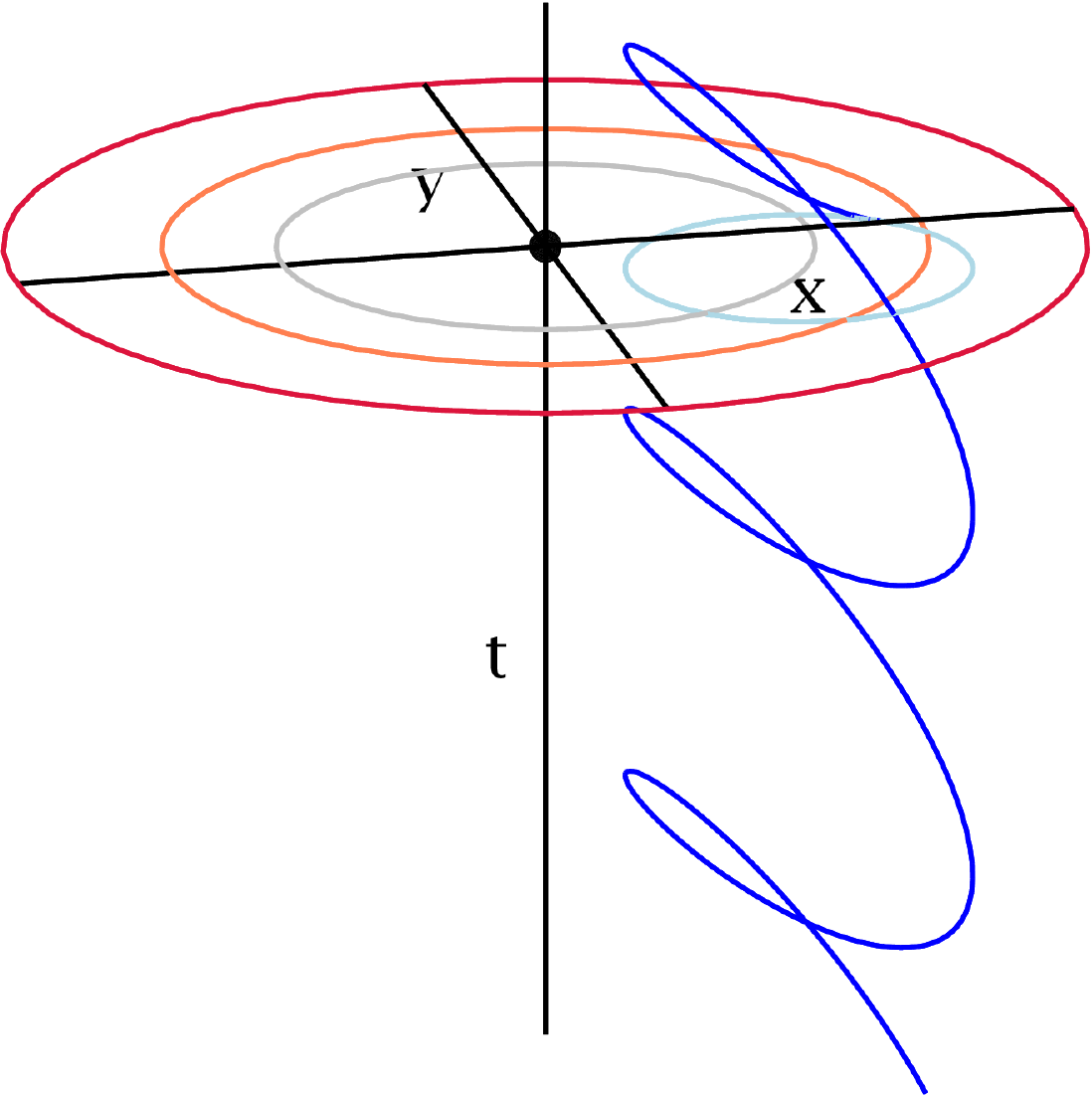}
	\caption{\centering \footnotesize $X$-$Y$-$t$-projection.}
\end{subfigure}%
\begin{subfigure}[H]{0.49\linewidth}
    \centering
	\includegraphics[width=0.99\textwidth]{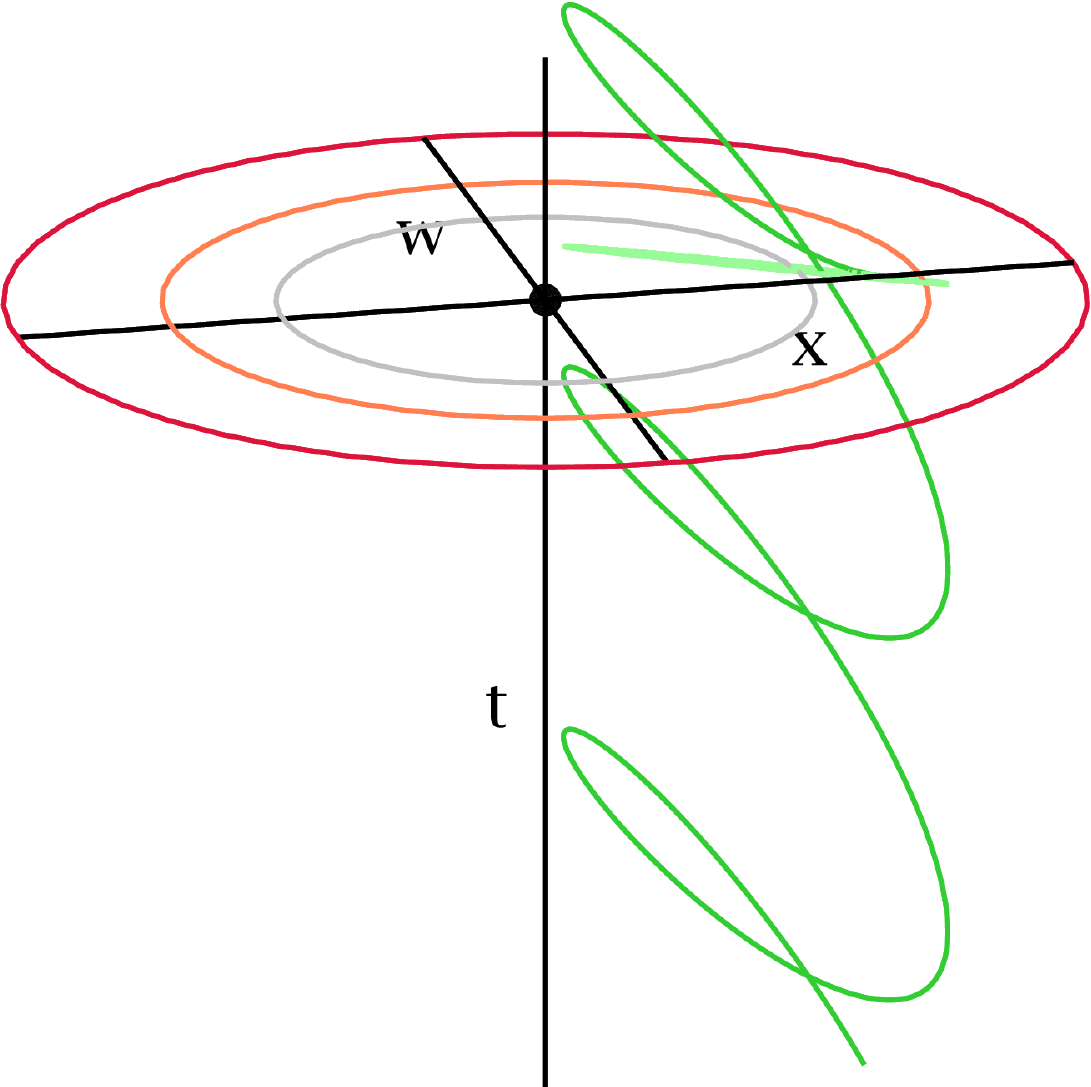}
	\caption{\centering \footnotesize $X$-$W$-$t$-projection.}
\end{subfigure}	
\caption{Bound orbits in different spatial-timelike projections: (a) and (b): Causal bound orbit (CBO) with $\delta =1, q=1.78,j=0.5,K=2,\Phi=-0.3,\Psi=0.1,E=10$. (c) and (d): Noncausal bound orbit (NBO) with $\delta =1, q=2,j=2.2,K=2,\Phi=-0.5,\Psi=0.1,E=5$.}\label{fig:caus_t}
\end{figure}

In order to obtain three-dimensional representations of the test particle motion, we simply omit one cartesian coordinate (e.g. the $W$-coordinate), which produces a projection of the orbital motion.\\
\\
In the following we present examples for bound orbits, especially its spatial and spatial-timelike projections. For the spatial-timelike projections we find Fig. \ref{fig:caus_t}, where the red circle describes the maximal radius of the test particle and the light blue/green one is the projection of the motion on the $X$-$Y$-plane with its time-evolution. Moreover the orange circle represents the modified and the grey one the classical G\"{o}del radius.

\begin{figure}[H]
\begin{subfigure}[H]{0.49\linewidth}
	\centering
	\includegraphics[width=0.99\textwidth]{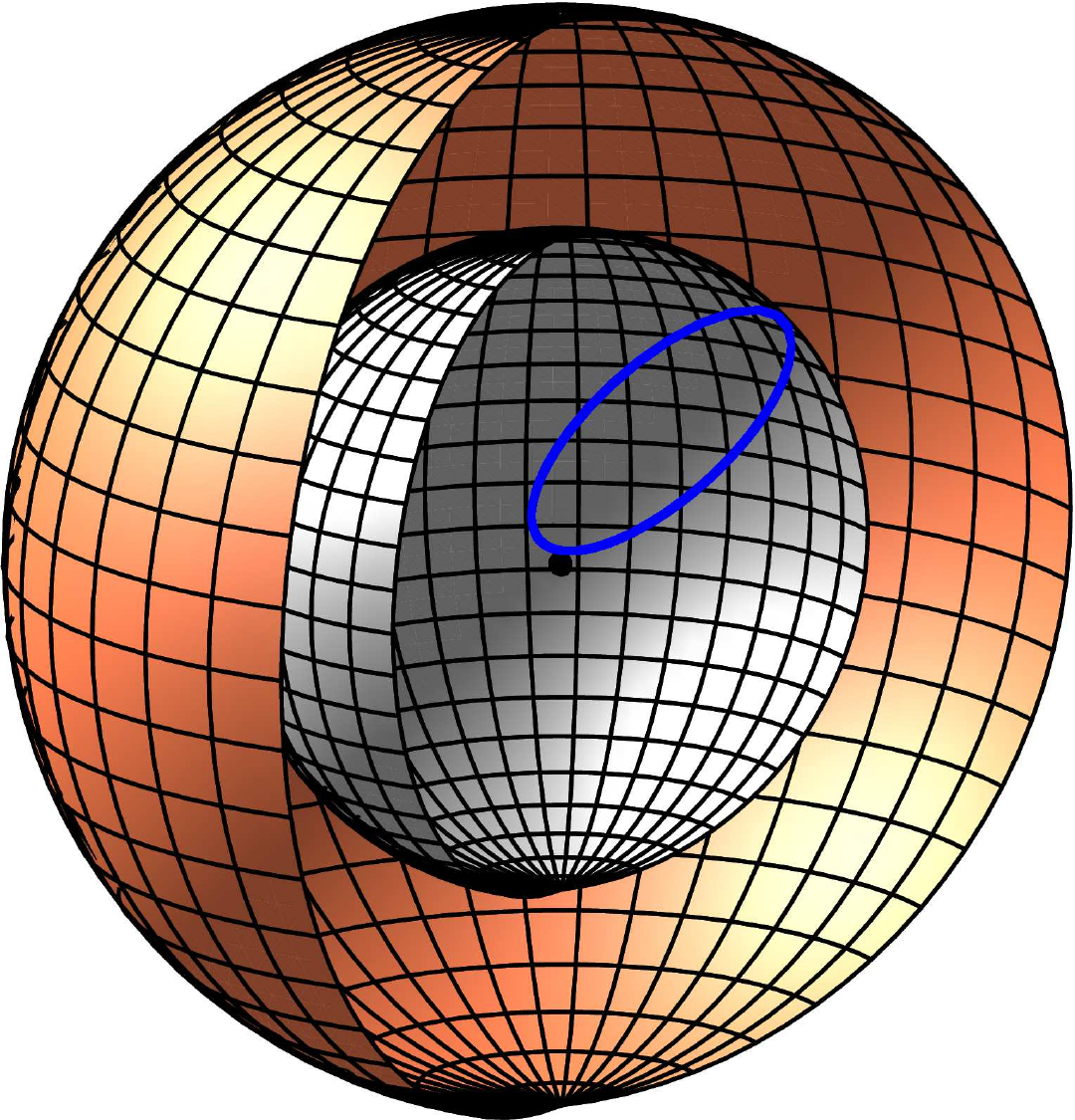}
	\caption{\centering \footnotesize $X$-$Y$-$Z$-projection.}
\end{subfigure}%
\begin{subfigure}[H]{0.49\linewidth}
	\centering
	\includegraphics[width=0.99\textwidth]{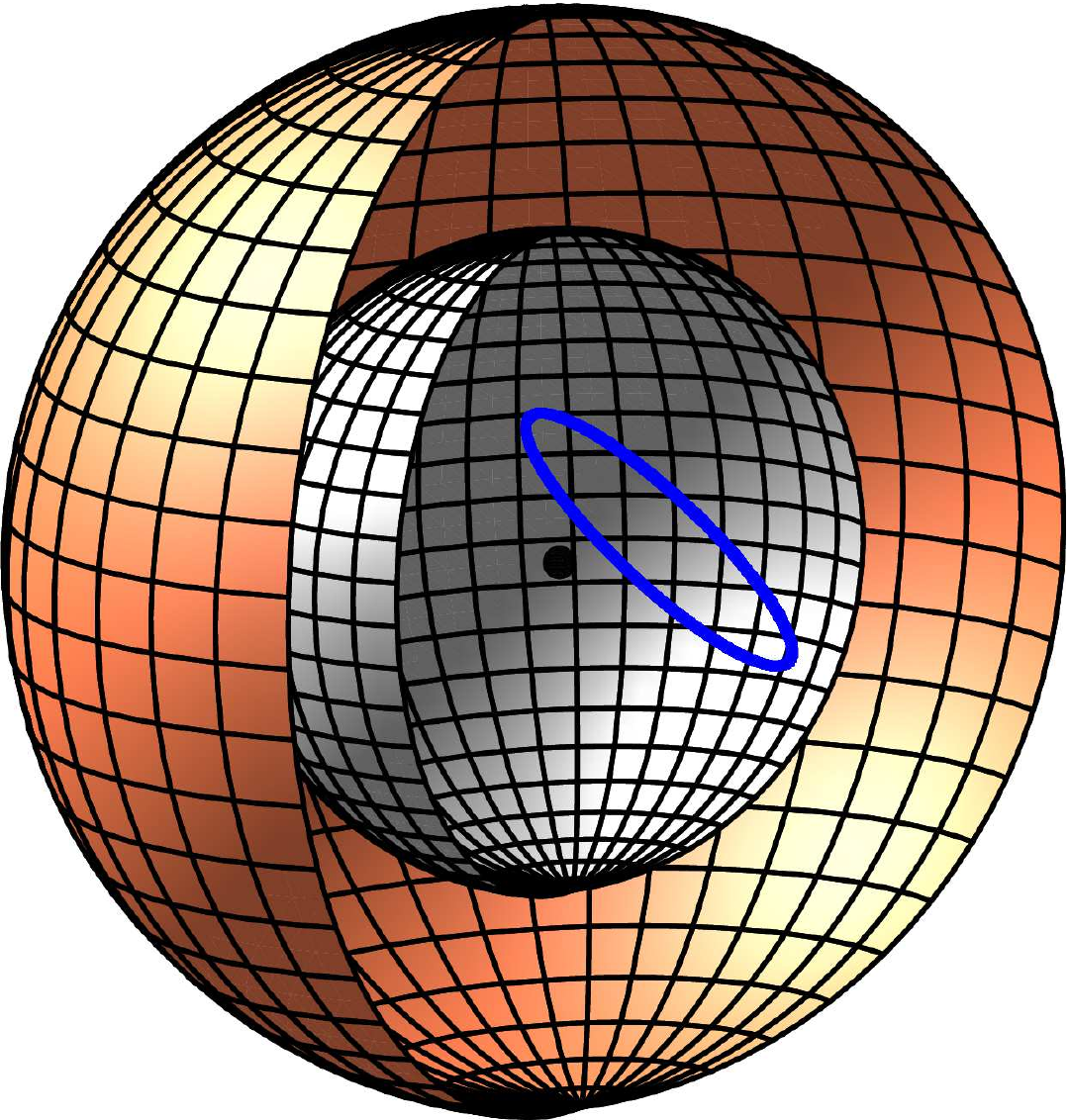}
	\caption{\centering \footnotesize $X$-$Y$-$W$-projection.}
\end{subfigure}\\[20pt]

\begin{subfigure}[H]{0.49\linewidth}
	\centering
	\includegraphics[width=0.99\textwidth]{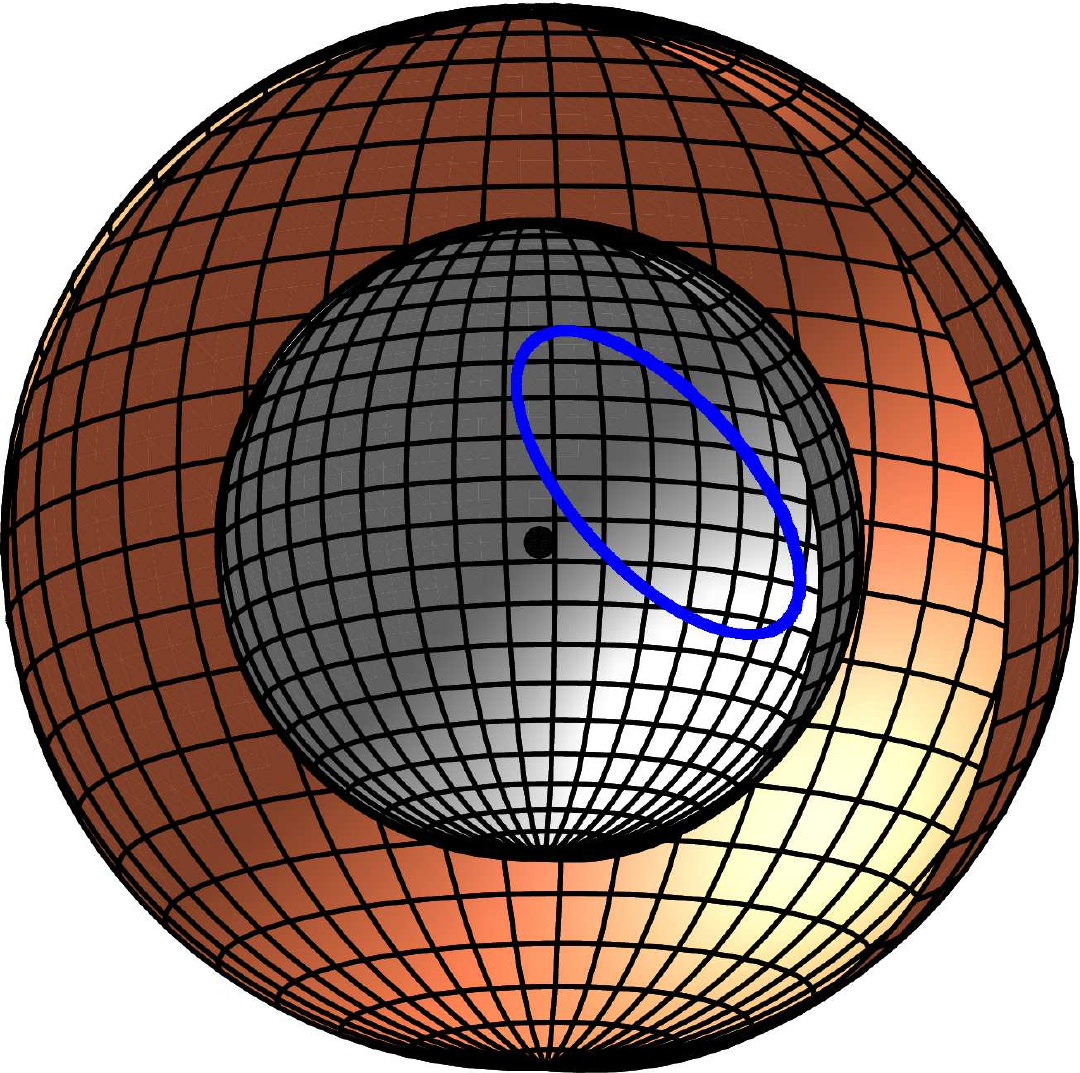}
	\caption{\centering \footnotesize $X$-$Z$-$W$-projection.}
\end{subfigure}%
\begin{subfigure}[H]{0.49\linewidth}
	\centering
	\includegraphics[width=0.99\textwidth]{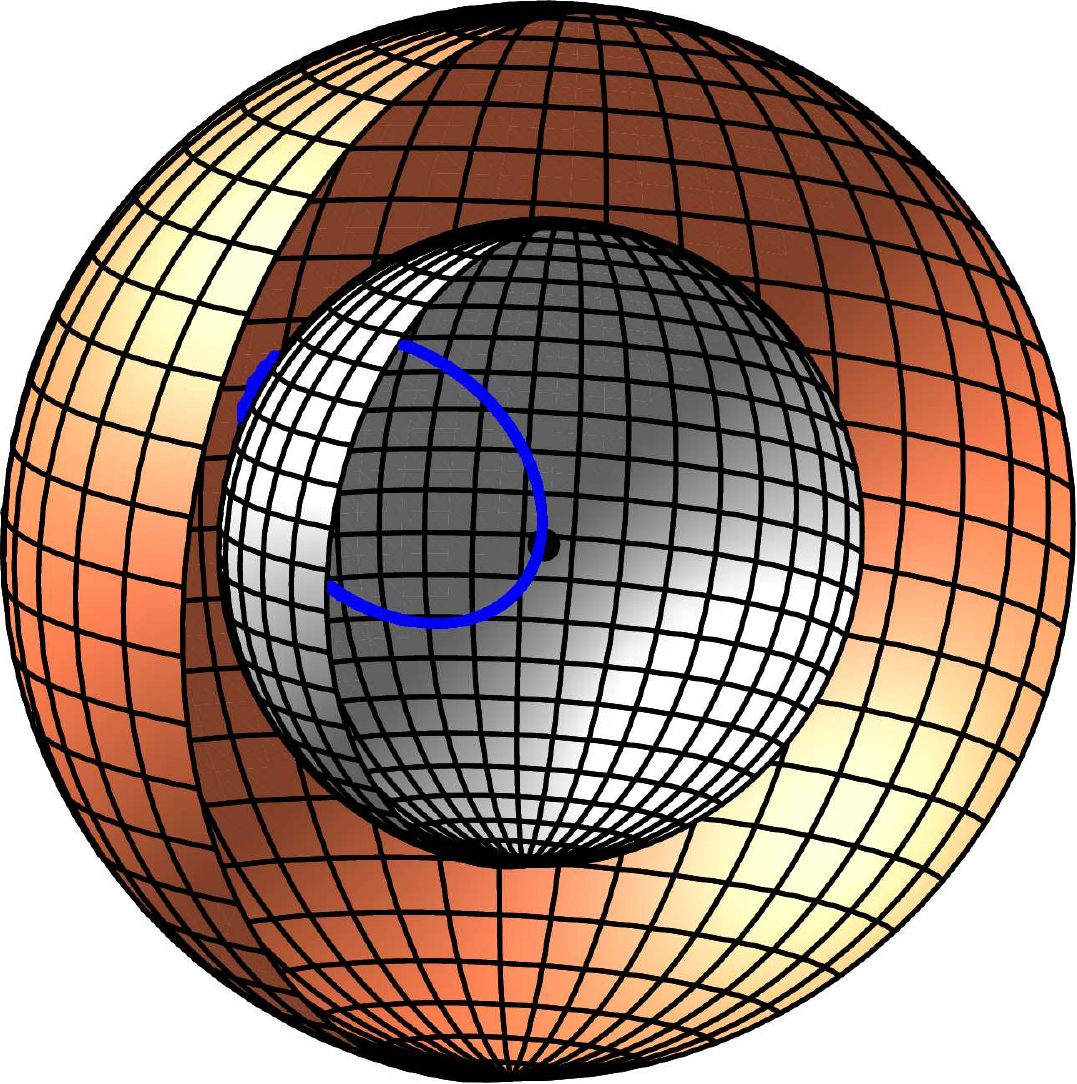}
	\caption{\centering \footnotesize $Y$-$Z$-$W$-projection.}
\end{subfigure}
	\caption{Causal bound orbits in different spatial projections using the parameter values $\delta =1, q=1.78,j=0.5,K=2,\Phi=-0.3,\Psi=0.1,E=10$.}  \label{fig:causal_orb}
\end{figure}

In Fig. \ref{fig:causal_orb} we present several spatial projections for a causal bound orbit. For the noncausal motion one finds similar results.

\subsection{CTGs}
\label{sec:CTG}

\begin{figure}[H]
\begin{subfigure}[H]{0.48\linewidth}
	\centering
	\includegraphics[width=0.99\textwidth]{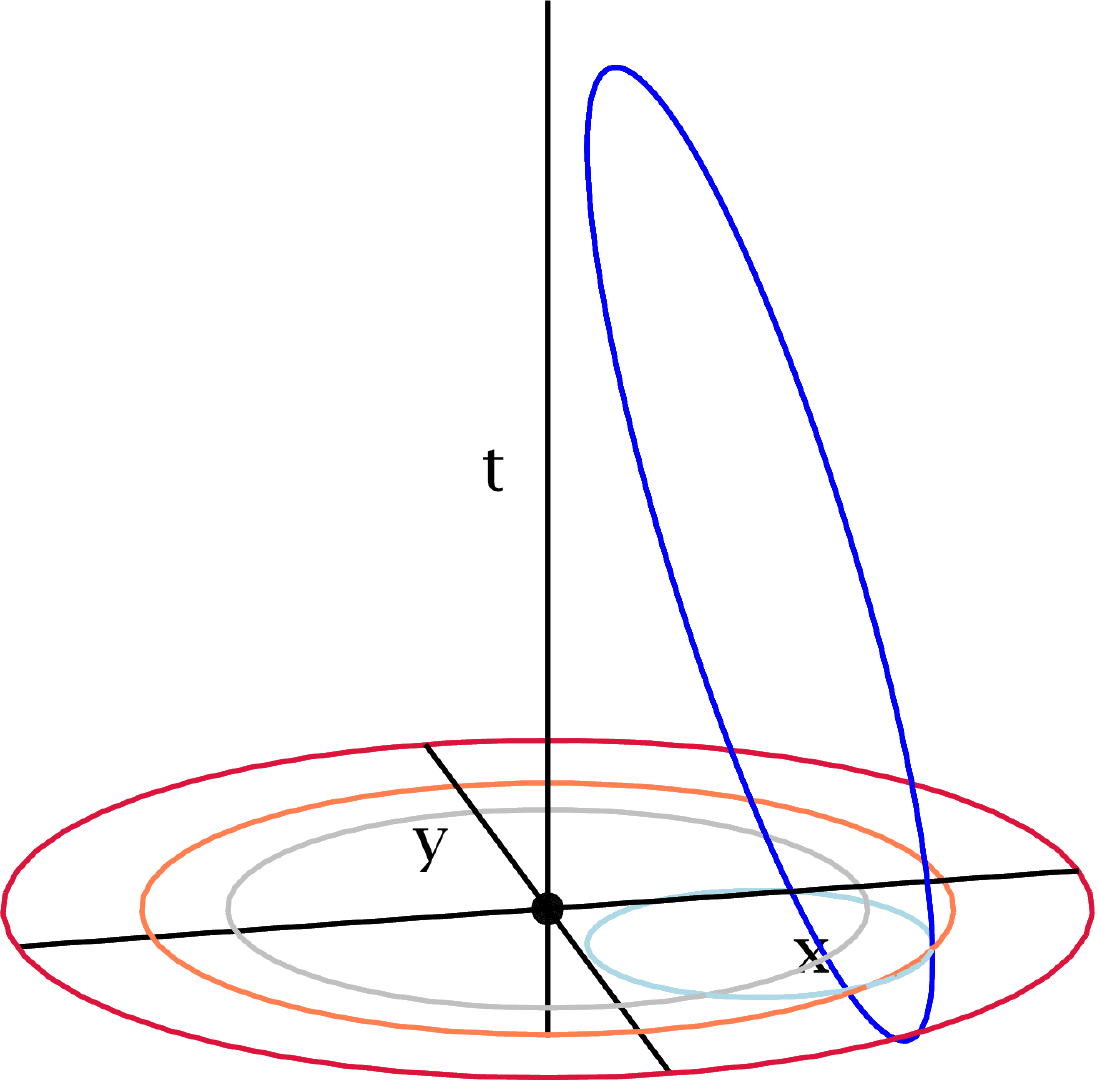}
	\caption{\centering \footnotesize $X$-$Y$-$t$-projection.}
\end{subfigure}%
\begin{subfigure}[H]{0.48\linewidth}
	\centering
	\includegraphics[width=0.99\textwidth]{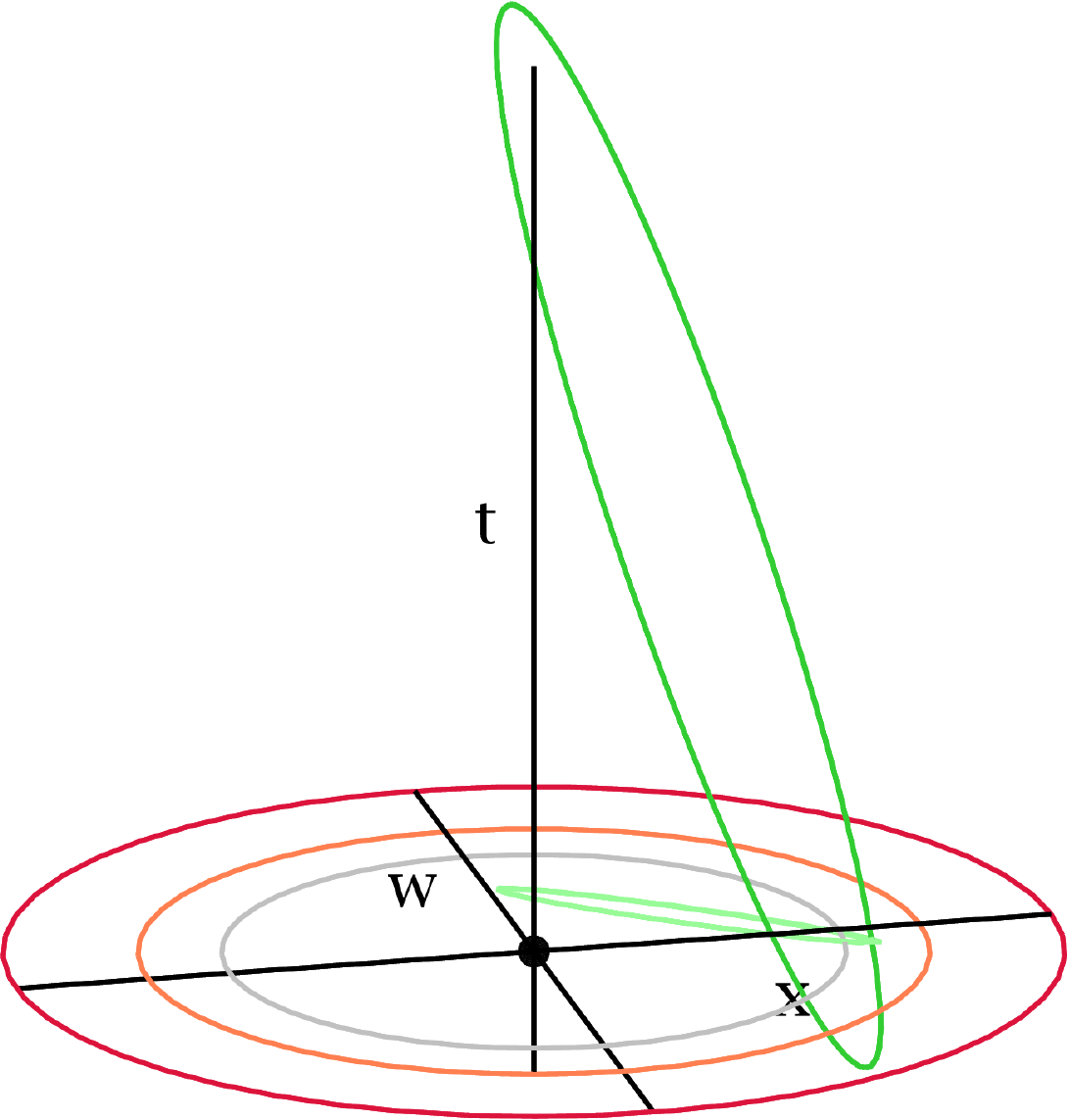}
	\caption{\centering \footnotesize $X$-$W$-$t$-projection.}
\end{subfigure}\\[20pt]

\centering
\begin{subfigure}[h]{0.48\linewidth}
	\centering
	\includegraphics[width=0.99\textwidth]{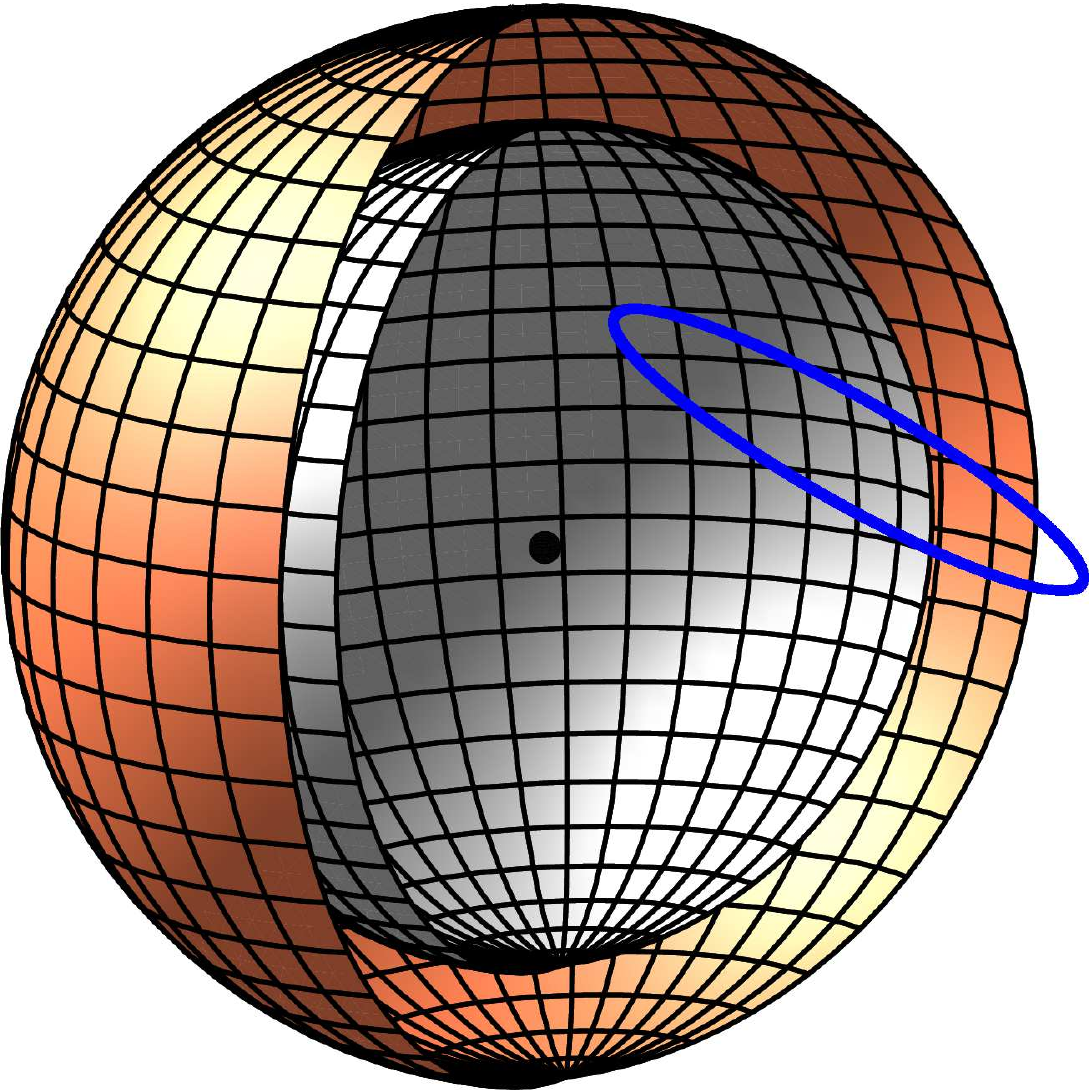}
	\caption{\centering \footnotesize $X$-$Y$-$Z$-projection.}
\end{subfigure}
	\caption{CTG orbits in different projections using the parameter values $\delta =1, q=1.78,j=0.5,K=2,\Phi=-0.3,\Psi=0.1,E=10$.}\label{fig:CTG_orb}
\end{figure}

As mentioned in \ref{sec:disc_motion}, CTGs can occur for a special choice of parameters. To obtain such a noncausal motion, we follow \cite{Gleiser:2005nw} and use the vanishing avarage of Eq.(\ref{eq:geot}) over one period which is given by

\begin{equation}
\bigg \langle \left( \frac{dt}{d \tau}  \right) \bigg \rangle_{_T} = (2 \sqrt{3}q - 4Ej^2) \langle r^4 \rangle_{_T} + (E-4 \Phi j) \langle r^2 \rangle_{_T} \overset{!}{=} 0,
\end{equation}

with

\begin{equation}
\langle r^n \rangle_{_T} = \frac{1}{T} \int_0^T r(\tau)^n d\tau
\end{equation}

and $T$ as the periodicity of $r(\tau)$.
Application of Eq.(\ref{eq:rsol}) and solving for $q$ leads to

\begin{equation}
q_{_{CTG}} = \frac{\sqrt{d_1 } + 4 \left(m + \frac{3}{4}  \right) E^2 - 8 j m \Phi E + \delta}{2 \sqrt{3}j(jE - 4\Phi m)}
\end{equation}
with $m=j^2-1$ and
\begin{align}
d_1 &= \left(8 m j^2 +1 \right)E^4 + 8\left(8m^2 \Phi^2 j^2 - \left(m + \frac{1}{2} \right)^2 \delta \right)E^2 \notag \\ & \quad +32 \left(m + \frac{1}{2} \right) j m \Phi E (\delta -E^2) + \delta^2.
\end{align}
Choosing this special $q_{_{CTG}}$, one obtains NBOs like the one in Fig.\ref{fig:CTG_orb}. The X-Y-t and X-W-t projections respectively show the closed time evolution and with this the character of a CTG.

As one can see in Fig.\ref{fig:caus_t} and Fig.\ref{fig:CTG_orb}, the maximal radius of the test particle for a noncausal motion exceeds the value of the modified G\"{o}del radius, obviously this holds also for a CTG.

\section{Conclusion and outlook}

In this paper we discussed the motion of massive and massless test particles in the five-dimensional G\"{o}del spacetime. We used the Hamilton-Jacobi formalism to derive the geodesic equations of motion and investigated their general properties. We also analyzed the effective potentials and studied the qualitative structure of the resulting orbits. According to this, we showed that the charge of the particle and the rotation parameter does not affect every equation of motion. Moreover we investigated the domain of the separation constant $K$ and found especially the restriction of $K \ge 0$. From the examination of causality, which could be found from the $t$ motion, we obtained relations between the energy $E$ and the two parameters $q$ and $j$, which must be satisfied for a causal motion. The geodesic equations were integrated analytically and the results were used to visualize the orbital motion. We showed that escape orbits are only possible for the special energy value of $E= \frac{\sqrt{3}}{2}q$. Consequently, there are no escape orbits in the case of (uncharged) lightlike motion.

As an outlook, one could think of calculating the orbits of charged test particles around higher-dimensional black holes in the G\"{o}del universe, which are coupled to the U(1) field.

\section*{Acknowledgement}

We gratefully acknowledge support by the Deutsche Forschungsgemeinschaft (DFG), in particular, within the framework of the DFG Research Training group 1620 {\it Models of gravity}.

\bibliographystyle{unsrt}

\begin{thebibliography}{99}
 
 %\cite{Godel:1949ga}
 \bibitem{Godel:1949ga} 
K.~G\"{o}del,
``An Example of a new type of cosmological solutions of Einstein's field equations of gravitation'',
Rev. Mod. Phys. {\bf 21}, 447 (1949).
 %doi:10.1103/RevModPhys.21.447
 
 
 %\cite{vanStockum:1937zz}
 \bibitem{vanStockum:1937zz} 
 W.~J.~van Stockum,
 ``The gravitational feild of a distribution of particles rotating about an axis of symmetry'',
 Proc.\ Roy.\ Soc.\ Edinburgh {\bf 57}, 135 (1937).
 
 
 %\cite{Kerr:1963ud}
 \bibitem{Kerr:1963ud} 
 R.~P.~Kerr, 
 ``Gravitational field of a spinning mass as an example of algebraically special metrics'',
 Phys. Rev. Lett. {\bf 11}, 237 (1963).
 
 %\cite{Gott:1990zr}
 \bibitem{Gott:1990zr} 
 J.~R.~Gott, III,
 ``Closed timelike curves produced by pairs of moving cosmic strings: Exact solutions'',
 Phys.\ Rev.\ Lett.\  {\bf 66}, 1126 (1991).
 %doi:10.1103/PhysRevLett.66.1126
 
 
 %\cite{Bondi:1996md}
 \bibitem{Bondi:1996md} 
 H.~Bondi and J.~Samuel,
 ``The Lense-Thirring effect and Mach's principle'',
 Phys.\ Lett.\ A {\bf 228}, 121 (1997).
 %doi:10.1016/S0375-9601(97)00117-5
 %[gr-qc/9607009].
 
 %\cite{Barbour:1995iu}
 \bibitem{Barbour:1995iu} 
 J.~B.~Barbour and H.~Pfister,
 ``Mach's principle: From Newton's bucket to quantum gravity. Proceedings, Conference, Tuebingen, Germany, July 26-30, 1993'',
 Boston, USA: Birkhaeuser (1995) 536 p. (Einstein studies. 6).
 
 
 %\cite{Hubble:1929ig}
 \bibitem{Hubble:1929ig} 
 E.~Hubble,
 ``A relation between distance and radial velocity among extra-galactic nebulae'',
 Proc.\ Nat.\ Acad.\ Sci.\  {\bf 15}, 168 (1929).
 %doi:10.1073/pnas.15.3.168
 
 
 %\cite{Hawking:1991nk}
 \bibitem{Hawking:1991nk} 
 S.~W.~Hawking,
 ``The Chronology protection conjecture'',
 Phys.\ Rev.\ D {\bf 46}, 603 (1992).
 %doi:10.1103/PhysRevD.46.603

 
 %\cite{Tangherlini:1963bw}
 \bibitem{Tangherlini:1963bw} 
 F.~R.~Tangherlini,
 ``Schwarzschild field in n dimensions and the dimensionality of space problem'',
 Nuovo Cim.\ {\bf 27}, 636 (1963).
 %doi:10.1007/BF02784569
 
 
 %\cite{Myers:1986un}
 \bibitem{Myers:1986un} 
 R.~C.~Myers and M.~J.~Perry,
 ``Black Holes in Higher Dimensional Space-Times'',
 Annals Phys. {\bf 172}, 304 (1986).
 
 
 %\cite{Gibbons:2004uw}
 \bibitem{Gibbons:2004uw} 
 G.~W.~Gibbons, H.~Lu, D.~N.~Page and C.~N.~Pope,
 ``The General Kerr-de Sitter metrics in all dimensions'',
 J.\ Geom.\ Phys.\ {\bf 53}, 49 (2005).
 %[hep-th/0404008].
 
 %\cite{Chen:2006xh}
 \bibitem{Chen:2006xh} 
 W.~Chen, H.~Lu and C.~N.~Pope
 ``General Kerr-NUT-AdS metrics in all dimensions'',
 Class.\ Quant.\ Grav.\ {\bf 23}, 5323 (2006).
 %[hep-th/0604125].
 
 
 %\cite{Emparan:2001wn}
 \bibitem{Emparan:2001wn} 
 R.~Emparan and H.~S.~Reall,
 ``A Rotating black ring solution in five-dimensions'',
 Phys.\ Rev.\ Lett.\ {\bf 88}, 101101 (2002).
 %doi:10.1103/PhysRevLett.88.101101
 %[hep-th/0110260].
 
 
  %\cite{Gauntlett:2002nw}
 \bibitem{Gauntlett:2002nw} 
 J.~P.~Gauntlett, J.~B.~Gutowski, C.~M.~Hull, S.~Pakis and H.~S.~Reall,
 ``All supersymmetric solutions of minimal supergravity in five dimensions'',
 Class. Quant. Grav. {\bf 20}, 4587 (2003).
% doi:10.1088/0264-9381/20/21/005
%[hep-th/0209114].


%\cite{Chong:2005hr}
\bibitem{Chong:2005hr} 
Z.~W.~Chong, M.~Cvetic, H.~Lu and C.~N.~Pope,
``General non-extremal rotating black holes in minimal five-dimensional gauged supergravity'',
Phys.\ Rev.\ Lett.\ {\bf 95}, 161301 (2005).
%[hep-th/0506029].


%\cite{Buser:2013uaa}
\bibitem{Buser:2013uaa} 
M.~Buser, E.~Kajari and W.~P.~Schleich,
``Visualization of the G\"odel universe'',
New J.\ Phys.\  {\bf 15}, 013063 (2013).
%doi:10.1088/1367-2630/15/1/013063
%[arXiv:1303.4651 [gr-qc]].



%\cite{Maldacena:1997re}
\bibitem{Maldacena:1997re} 
J. M. Maldacena, \glqq The Large N limit of superconformal field theories and supergravity\grqq,
Int. J. Theor. Phys. {\bf 38}, 1113 (1999).
%[Adv.\ Theor.\ Math.\ Phys.\  {\bf 2}, 231 (1998)]
%doi:10.1023/A:1026654312961
%[hep-th/9711200].


%\cite{Grave:2009zz}
\bibitem{Grave:2009zz} 
F.Grave, M.~Buser, T.~Muller, G.~Wunner and W.~P.~Schleich,
``The Godel universe: Exact geometrical optics and analytical investigations on motion'',
Phys.\ Rev.\ D {\bf 80}, 103002 (2009).
%doi:10.1103/PhysRevD.80.103002
%%CITATION = doi:10.1103/PhysRevD.80.103002;%%
%5 citations counted in INSPIRE as of 19 Nov 2017


%\cite{Reimers:2016czc}
\bibitem{Reimers:2016czc} 
S.~Paranjape and S.~Reimers,
``Dynamics of test particles in the five-dimensional, charged, rotating Einstein-Maxwell-Chern-Simons spacetime'',
Phys.\ Rev.\ D {\bf 94}, no. 12, 124003 (2016).
%doi:10.1103/PhysRevD.94.124003
%[arXiv:1609.03557 [gr-qc]].


%\cite{Gimon:2003ms}
\bibitem{Gimon:2003ms} 
E.~G.~Gimon and A.~Hashimoto,
``Black holes in Godel universes and pp waves'',
Phys. Rev. Lett. {\bf 91}, 021601 (2003).
%doi:10.1103/PhysRevLett.91.021601
%[hep-th/0304181].


%\cite{Wu:2007gg}
\bibitem{Wu:2007gg} 
S.~Q.~Wu,
``General Non-extremal Rotating Charged Godel Black Holes in Minimal Five-Dimensional Gauged Supergravity'',
Phys.\ Rev.\ Lett.\  {\bf 100}, 121301 (2008).
%[arXiv:0709.1749 [hep-th]].


%\cite{Grunau:2017uzf}
%\bibitem{Grunau:2017uzf} 
%S.~Grunau, H.~Neumann and S.~Reimers,
%``Geodesic motion in the five-dimensional Myers-Perry-AdS spacetime'',
%arXiv:1711.02933 [gr-qc].
%%CITATION = ARXIV:1711.02933;%%


%\cite{Misner:1974qy}
\bibitem{Misner:1974qy} 
C.~W.~Misner, K.~S.~Thorne and J.~A.~Wheeler,
``Gravitation'',
 W. H. Freeman and Company, San Francisco (1973).


%\cite{Mino:2003yg}
\bibitem{Mino:2003yg} 
Y.~Mino,
``Perturbative approach to an orbital evolution around a supermassive black hole'',
Phys.\ Rev.\ D {\bf 67}, 084027 (2003).
  %[gr-qc/0302075].


%\cite{Gradshteyn:2007}
\bibitem{Gradshteyn:2007}
I.~S.~Gradshteyn and I.~M.~Ryzhik,
``Table of Integrals, Series, and Products'' (Academic Press, Seventh Edition, 2007).

%\cite{Gleiser:2005nw}
\bibitem{Gleiser:2005nw} 
R.~J.~Gleiser, M.~Gurses, A.~Karasu and O.~Sarioglu,
``Closed timelike curves and geodesics of G\"{o}del-type metrics'',
Class.\ Quant.\ Grav.\  {\bf 23} (2006) 2653-2664.
%[arXiv:gr-qc/0512037].













%%\cite{Herdeiro:2002ft}
%\bibitem{Herdeiro:2002ft} 
%C.~A.~R.~Herdeiro,
%``Spinning deformations of the D1 - D5 system and a geometric resolution of closed timelike curves'',
%Nucl. Phys. B {\bf 665}, 189 (2003).
%%doi:10.1016/S0550-3213(03)00484-X
%%[hep-th/0212002].
%
%
%
%%\cite{Kunduri:2005vc}
%\bibitem{Kunduri:2005vc} 
%H.~K.~Kunduri and J.~Lucietti,
%``Three charge supertubes in type IIB plane wave backgrounds'',
%JHEP {\bf 0509}, 014 (2005).
%%doi:10.1088/1126-6708/2005/09/014
%%[hep-th/0506222].



 \end{thebibliography}

\end{document}